\documentclass[journal]{IEEEtran}

\usepackage{amsmath,amsthm}
\usepackage{amsfonts,amssymb}
\usepackage{CJK}
\usepackage{multicol}
\usepackage{multirow}
\usepackage{diagbox}
\usepackage{graphicx}
\usepackage{subfig} %Á½¸öͼƬ²¢ÁдæÔÚ
\usepackage{bm} %¼Ó´ÖÌØÊâ·ûºÅ
\usepackage{lineno} %ÏÔʾÐкÅ
\allowdisplaybreaks % allow the equation display in different pages
\usepackage{epstopdf} % transfer the figure with eps formation to figure with pdf formation
\usepackage{pgf,tikz}
\usepackage{tikz}
\usetikzlibrary{arrows,shapes,snakes,shadows,positioning,automata,patterns}
\usetikzlibrary{trees,decorations.pathmorphing,decorations.markings}
\usetikzlibrary{backgrounds}

\usepackage[%dvipdfm,  %pdflatex,pdftex % ÕâÀï¾ö¶¨ÔËÐÐÎļþµÄ·½Ê½²»Í¬
            pdfstartview=FitH,
            CJKbookmarks=true,
            bookmarksnumbered=true,
            bookmarksopen=true,
            colorlinks, %×¢Ê͵ô´ËÏîÔò½»²æÒýÓÃΪ²ÊÉ«±ß¿ò(½«colorlinksºÍpdfborder ͬʱעÊ͵ô)
            pdfborder=001,   %×¢Ê͵ô´ËÏîÔò½»²æÒýÓÃΪ²ÊÉ«±ß¿ò
            linkcolor=blue,
            anchorcolor=blue,
            citecolor=blue
            ]{hyperref}

\usepackage{authblk}
\usepackage{booktabs}

\newtheorem{theorem}{Theorem}
\newtheorem{assumption}{Assumption}

\newtheorem{lemma}{Lemma}

\newtheorem{definition}{Definition}

\newtheorem{remark}{Remark}

\begin{document}
%\linenumbers %¿ªÊ¼ÏÔʾÐкÅ

\title{Distributed Nash Equilibrium Seeking Algorithm Design for Multi-Cluster Games with High-Order Players}

%\author{Zhenhua~Deng
%        and~Yiguang~Hong
%\thanks{This work was supported by the National Key Research and Development Program of China (2016YFB0901902) and NSFC (61733018, 61333001, 61573344).}
%\thanks{Z. Deng is with the School of Information Science and Engineering, Central South University, Changsha, 410004, China (e-mail: zhdeng@amss.ac.cn).}
%\thanks{S. Liang is with the School of Automation and Electrical Engineering, University of Science and Technology Beijing, Beijing, 100083, China (e-mail: sliang@amss.ac.cn).}
%\thanks{Y. Hong is with the Key Lab of Systems and Control, Academy of Mathematics and System Science, Chinese Academy of Sciences, Beijing, 100190, China (yghong@iss.ac.cn).}}

\author{Zhenhua Deng, Yangyang Liu
        %Yiguang Hong,~\IEEEmembership{Fellow,~IEEE,}
\thanks{This work was supported by  the National Natural Science Foundation
of China under Grants 61803385 and 31870532, the Hunan Provincial Natural Science
Foundation of China under Grants 2019JJ50754.}
\thanks{Z. Deng and Y. Liu are with  the School of Automation, Central South University, Changsha, 410075, China
(e-mail: zhdeng@amss.ac.cn; yyangliu@csu.edu.cn).}% <-this % stops a space
}

\maketitle

\begin{abstract}
In this paper, a multi-cluster game with high-order players is investigated.
Different from the well-known multi-cluster games, the dynamics of players are taken into account in our problem.
Due to the high-order dynamics of players, existing algorithms for multi-cluster games cannot solve the problem.
For purpose of seeking the Nash equilibrium of the game, we design a distributed algorithm based on gradient descent and state feedback, where a distributed estimator is embedded for the players to estimate the decisions of other players.
Furthermore, we analyze the exponential convergence of the algorithm via variational analysis and Lyapunov stability theory.
Finally, a numerical simulation verifies the effectiveness of our method.
\end{abstract}


\IEEEpeerreviewmaketitle

\section{Introduction}
Distributed optimization and noncooperative games describe cooperative and competitive behaviors among multiple agents, respectively, which have widely applications in a variety of fields, such as smart grids, social networks, parameter estimation and radio networks, and have attracted considerable attention (see \cite{Gharesifard2016Price,ghaderi2014opinion,Ram2010ParameterEstimation,cao2020,Yuan2021,Lou16}).
%such as cognitive networks, robot networks and smart grids. %\cite{Saraydar2002Efficientpowercontro;l}).
%\cite{Qu2018}%ÓÅ»¯ÎÄÕ£¬ÔÝʱȥµô
%which have received much attention.
%Recently, researchers have made great efforts in the algorithm design for distributed optimization (see \cite{deng2018resource}, \cite{Lin17}, \cite{Deng16}, \cite{Nedic2009}, \cite{zeng2017optimization}, \cite{Qu2018}, \cite{Li2020}) and noncooperative games problems (see \cite{deng2018distributed}, \cite{Saraydar2002Efficient power control}, \cite{Li2020Femtocell Networks}, \cite{SALEHISADAGHIANI201927}, \cite{ deng2018distributed}, \cite{Lu2019, deng2018distributeddisturbed}, \cite{deng2019distributedEL}, \cite{D.Gadjov2019noncoop-game}).
%%%%%%%%%%%%%%%%%%%%%%%%%%%%%%%%%%%%%%%%%%%%%%%
%Distributed optimization problems have widely applications in a variety of fields, for instance, smart grids, parameter estimation and cognitive networks, in which participants cooperatively optimize the the sum of the cost functions of all agents (see \cite{Gharesifard2016Price}, \cite{Ram2010ParameterEstimation}, \cite{Lorenzo12}).
%%\cite{projection,liang2018distributed,lakshmanan2008decentralized}, \cite{bretthauer1997quadratic}, \cite{Xiao06}).
%%\cite{Lin17}, \cite{Nedic2009},  \cite{Li2020}),
%On the other hand, every player competitively minimizes its own cost function in noncooperative games, which have been broadly applied in social networks, power systems and radio networks (see \cite{ghaderi2014opinion}, \cite{hobbs2007nash}, \cite{scutari2014real}).
%%%%%%%%%%%%%%%%%%%%%%%%%%%%%%%%%%%%%%%%%%%%%%%%%%%
In distributed optimization problems, all participants cooperate with their neighbors to search the optimal solution of the networks (see \cite{liang2019exponential,yang2017distributed,liy2019,he2019continuous,kia2015distributed,Zhang17}),
%\cite{projection,liang2018distributed,lakshmanan2008decentralized}, \cite{bretthauer1997quadratic}, \cite{Xiao06}).
%\cite{Lin17}, \cite{Nedic2009},  \cite{Li2020}),
while in noncooperative games, every player competes with other players to  selfishly minimize its own cost function (see \cite{ZhangLWJ19,deng2021distributedEL,bianchiG2021,RomanoP20}).
%{deng2018distributed}, \cite{deng2019distributedEL}, \cite{D.Gadjov2019noncoop-game}, \cite{C.De2020noncoop-game}).
Nevertheless, it is noteworthy that in numerous engineering practices, cooperation and competition among agents always coexist, such as healthcare networks and transportation networks (see \cite{shehory1998methods}, \cite{peng2009coexistence}).
%For these cases, separate noncooperative games or distributed optimization methods may not
%In order to the coexistence of and competition among the participants,
Multi-cluster games can simultaneously characterize cooperation relationship within clusters and competition relationship between clusters, which extends the aforementioned distributed optimization problems and noncooperative game problems, and have aroused the interest of many scholars (see \cite{YE2018266},\cite{ZENG201920}, \cite{Ye2019AUnifiedStrategy}, \cite{meng2020linear}, \cite{pang2020gradient}).

Multi-cluster games are conducted by multiple interacting clusters, and each cluster consists of a group of players.
Every cluster wants to minimize its own cost function that is a summation of the cost functions of all players in the cluster.
%, whereas players in the same cluster collaboratively minimize the cost function of this cluster.
Consequently, the objective of these clusters are to seek the Nash equilibrium of the multi-cluster games.
 %there is a and each cluster is composed of a group of players, which
%(remark/conclusion)With the proposed algorithm, it is not necessary for the players to know the decisions of players in the same cluster, nor is it necessary to know the decisions of players in any other clusters.
To this end, some Nash equilibrium seeking algorithms have been proposed for multi-cluster games recently.
For example, for unconstrained multi-cluster games, \cite{YE2018266}
%firstly formulated the problem of  N-coalition noncooperative games and
designed a Nash equilibrium seeking algorithm based on dynamic average consensus, and for constrained multi-cluster games, \cite{ZENG201920} presented a distributed Nash equilibrium seeking algorithm via projected operators.
In order to reduce the communication and computation costs, \cite{Ye2019AUnifiedStrategy} exploited a Nash equilibrium seeking algorithm for multi-cluster games with interference graphs.
%In order to reduce the communication and computation costs, \cite{Ye2019AUnifiedStrategy} presented a NE seeking algorithm by utilizing interference graphs.
For multi-cluster games with partial-decision information, \cite{meng2020linear} proposed a distributed Nash equilibrium seeking algorithm based on the intra- and inter-communication of clusters.
%The decisions of players are limited to multifarious constraints in plenty of engineering applications, which are not concerned in the aforementioned results.
For multi-cluster games with non-smooth cost functions, \cite{pang2020gradient} developed a Nash equilibrium seeking algorithm by Gaussian smoothing techniques.

%Moreover, for multi-cluster games with unknown cost functions, %\cite{YE2020108815} designed an extremum seeker-based approach and
%\cite{pang2020nash} designed a gradient-free NE seeking algorithm by combining Gaussian smoothing techniques and gradient tracking.

Cyber-physical systems (CPSs) integrate computation, communication and physical processes, and commonly appear in multifarious engineering applications, such as power networks and transportation systems (see \cite{Kim12cyber}, \cite{Zhang18}).
%due to its society, economy and environment benefits \cite{Kim12}.
%Remarkably, there are many distributed optimization problems in CPSs, such as the optimal power flow of smart grids and the unknown source search of multirobot (see \cite{Zhang18,Zhang17}).
%In the distributed optimization of CPSs, physical systems can be used to perform distributed optimization tasks (e.g., the unknown source search of multirobot \cite{Zhang17}).
With the development of CPSs, more and more distributed algorithms involved with the dynamics of systems have been exploited to study how physical systems autonomously accomplish distributed tasks.
%for distributed optimization problems and noncooperative games problems. (see \cite{deng2019distributedEL}, \cite{Zhang18}, \cite{dengliang2018distributed}, \cite{deng2019distributed}).
For example, \cite{ZhangLWJ19,deng2021distributedEL} investigated aggregative game problems of disturbed systems and  Euler-Lagrange systems, respectively, and \cite{Zhang17,deng2019distributedre} studied distributed optimization problems of second-order systems.
On the other hand, many physical systems, such as generators, robots and satellites, can be depicted by  high-order systems, and first- and second-order systems can be viewed as the special cases of high-order systems.
Nevertheless, to the best of our knowledge, there are no results about multi-cluster games with high-order multi-agent systems.
Moreover, without further integrating the control of high-order dynamics, existing Nash seeking algorithms for multi-cluster games, such as \cite{YE2018266}, \cite{ZENG201920}, \cite{Ye2019AUnifiedStrategy}, \cite{meng2020linear}, \cite{pang2020gradient}, are ineffective for the problem. These observations motivate us to study multi-cluster games of high-order multi-agent systems.

The objective of this paper is to investigate multi-cluster games of high-order players and design a distributed algorithm to seek the Nash equilibrium of the game.
The contributions of this paper are summarized as follows:
\begin{enumerate}
\item We study the multi-cluster games of multi-agent systems, where the players have high-order dynamics.
The formulation extends non-cooperative games discussed in \cite{ZhangLWJ19,deng2021distributedEL,bianchiG2021,RomanoP20}
%\cite{deng2018distributed}, \cite{deng2019distributedEL}, \cite{D.Gadjov2019noncoop-game}, \cite{C.De2020noncoop-game}
by containing the distributed optimization of players within clusters,
the distributed optimization problems studied in \cite{liang2019exponential,yang2017distributed,liy2019,he2019continuous,kia2015distributed,Zhang17}
%\cite{liang2018distributed}, \cite{lakshmanan2008decentralized}, \cite{bretthauer1997quadratic}, \cite{Xiao06}
    %\cite{deng2020resourceHighOrder}, \cite{Xiao06}, \cite{L.Jin2018resource}, \cite{R.Shang2016resourcr}, \cite{M.Zargham2014resource}
    by considering the noncooperative games between clusters,
and the multi-cluster games investigated in \cite{YE2018266,ZENG201920,Ye2019AUnifiedStrategy,meng2020linear,pang2020gradient} by adding the high-order dynamics of players.
    Because the players have high-order dynamics, existing algorithms for multi-cluster games, such as \cite{YE2018266,ZENG201920,Ye2019AUnifiedStrategy,meng2020linear,pang2020gradient}, cannot be applied to our problem.
    %the high-order dynamics are considered for the players, which are more complex than the cases without CPSs in a majority of results (see \cite{YE2018266}, \cite{Ye2019AUnifiedStrategy}, \cite{meng2020linear}, \cite{pang2020gradient}, \cite{ZENG201920}).

%Without involving the high-order dynamics, existing distributed resource allocation algorithms, such as  \cite{Xiao06,Lakshmanan08,Beck14,Wei13,LiLYCL,Cherukuri15,ZengYHX,Liang18,Deng19,DengLH17,Yu17,DengLY18,Deng}, cannot control high-order agents to autonomously accomplish  resource allocation tasks.
%Moreover, the high-order dynamics of agents and the nonlinearity of cost functions lead to the difficulty in algorithm design and analysis.
\item We design a distributed algorithm based on state feedback and gradient descent to seek the Nash equilibrium of multi-cluster games.
%    Specifically, projection operations are used to settle the local set constraints, and subgradients and differential inclusions are able to deal with the nonsmooth cost functions.
%A distributed estimator is embedded in the algorithm for the estimation of the decisions of players.
Most of  existing multi-cluster game algorithms need full decision information of all players (see
\cite{YE2018266}, \cite{Ye2019AUnifiedStrategy}, \cite{pang2020gradient}), and in contrast, the
players with our algorithm only exchange information with their neighbors.
%in the same cluater, and their cost functions and gradients are not shared with any other players.
Furthermore, we analyze the convergence of the algorithm via variational analysis and Lyapunov stability theory. Compared with the algorithms in \cite{ZhangLWJ19,deng2021distributedEL,bianchiG2021,RomanoP20,ZENG201920}, our algorithm exponentially instead of asymptotically converges to the Nash equilibrium of  multi-cluster games .
\end{enumerate}

The paper is organized as follows.
Section \ref{sec.basis} introduces some basic knowledge and describes our problem.
Section \ref{sec.r} presents a distributed Nash equilibrium seeking algorithm and analyzes its convergence.
Section \ref{sec.s}  provides a numerical example to illustrate the algorithm.
Finally, Section \ref{sec.conclusion} summarizes the conclusion.

%$\mathbb{R}$ and $\mathbb{N}$ are the sets of real and natural numbers, respectively.
%$\mathbb{R}_+$ and $\mathbb{R}_-$ are the sets of positive and negative real numbers, respectively.
%Specifically,
%For a vector $x \in \mathbb{R}^{n} $, $\left\|  x \right\|_1=\sum_{i=1}^n |x_i|$, $\left\|  x \right\|_2=\sqrt{x^T x}$, and $\left\|  x \right\|_\infty=$max$_{i=1,\dots,n}\{|x_i|\}$.

\textsl{Notations:}
$\mathbb{R}$ and $\mathbb{R}^{n} $ represent the set of real numbers and the $n$-dimensional Euclidean space, respectively.
$\otimes$ is the Kronecker product.
$\times$ is the Cartesian product.
${0_n}$ and ${1_n}$ denote the column vectors of $n$ zeros and ones, respectively.
$\left\| x  \right\|$ is the standard Euclidean norm of vector $x$.
$\| A\|$ is the spectral norm of matrix $A$.
Define $col(x_1,...,x_n)=[x_1^T, ..., x_n^T]^T$, where $x_i$ is a vector. % or the $i$th element of vector $x$.
${I_n}$ denotes the identity matrix.
%$col(\textbf{x}_1,...,\textbf{x}_n)=[\textbf{x}_1^T, ..., \textbf{x}_n^T]^T$  $\textbf{x}_i$
Let $\lambda_{min}(A)$ and $\lambda_{max}(A)$ be the smallest and the largest eigenvalues of matrix $A$, respectively.

\section{Preliminaries and Formulation}\label{sec.basis}

In this section, some preliminaries about graph theory and variational analysis are reviewed, and then our problem is formulated.

\subsection{Preliminaries}
Here some concepts about graph theory are presented (see \cite{Godsil01graph}).
Consider an undirected graph $\mathcal{G}:=\{\mathcal{V}, \mathcal{E}, \mathcal{A}\}$, where $\mathcal{V}= \{ 1, \ldots, N\}$ is the vertex set, $\mathcal{E} \subseteq \mathcal{V} \times \mathcal{V}$ is the edge set, and $\mathcal{A}=[a_{ij}]_{N \times N}$ is the adjacency matrix.
%and $\mathcal{A} = {\left[ {{a_{ij}}} \right]_{N \times N}}$ is the adjacency matrix with ${a_{ij}}$ being the weighting of edge $(i,j)$.
An edge of $\mathcal{G}$ is denoted by $\{i,j\} \in \mathcal{E}$ if vertexes $i$ and $j$ can receive information from each other, i.e., they are neighbors.
Besides, $\{i,i\} \notin \mathcal{E}$, $\forall \, i \in \mathcal{V}$.
%If there are a sequence of distinct vertexes ${i_0}, {i_1}, \ldots, {i_l}$ such that ${i_k}$ and $i_{k+1}$ are neighbors, $\forall \, k \in \{0,1,\ldots, l-1\}$, then these vertexes form a path of $\mathcal{G}$.
A path of $\mathcal{G}$ is given by a sequence of distinct vertexes connected by edges.
%A graph is undirected if and only if ${a_{ij}} = {a_{ji}}$ for all $i,j \in \mathcal{V}$.
The undirected graph $\mathcal{G}$ is connected if there exists a path between any pair of vertexes.
%Denote $\mathcal{A} = {\left[ {{a_{ij}}} \right]_{N \times N}}$ as the weighted adjacency matrix of $\mathcal{G}$ with ${a_{ij}}$ being the weighting of $\{i,j\}$, where ${a_{ij}} = {a_{ji}} > 0$ if $\{i,j\} \in \mathcal{E}$, and ${a_{ij}} = 0$, otherwise.
The element $a_{ij}$ of the adjacency matrix $\mathcal{A}$ represents the weighting of $\{i,j\}$, where ${a_{ij}} = {a_{ji}} > 0$ if $\{i,j\} \in \mathcal{E}$, and ${a_{ij}} = 0$, otherwise.
Let $\mathrm{deg}_{i} = \sum_{j = 1}^N {{a_{ij}}} $ be the degree of vertex $i$.
%In this case, ${i_0}$ and ${i_l}$ are called the end nodes of the path.
%The undirected graph is connected if, for every pair of nodes, there is a path that has the two nodes as its end nodes.
Define $L=\mathcal{D}-\mathcal{A}$ as the Laplacian matrix of $\mathcal{G}$, where $\mathcal{D} = diag\{\mathrm{deg}_1, \ldots \mathrm{deg}_N\} $.
Obviously, $L{1_N} = 0_N$. % and ${1_N^T }L = 0_N^T$.
%The graph is weight-balanced if and only if $1_N^T\mathcal{L} = 0$.
The eigenvalues of $L$ are expressed as ${\lambda _1}, \ldots {\lambda _N}$, where ${\lambda _i} \le {\lambda _j}$ if $i \le j$.
%If the graph is weigh-balanced, ${\lambda _1} = 0$.
Moreover, $\mathcal{G}$ is connected if and only if ${\lambda _2} > 0$.

Next, some definitions about variational analysis are introduced (see \cite{facchinei2003finite}).

A function $f: {\mathbb{R}^n} \to \mathbb{R}$ is convex if for any $  x, y \in {\mathbb{R}^n}$,
\begin{equation*}
 f(\alpha x + (1-\alpha)y) \le \alpha f(x) + (1-\alpha)f(y), ~ \forall \, \alpha \in [0,~ 1].
\end{equation*}
%A differentiable function $f:{\mathbb{R}^n} \to \mathbb{R}$ is  $\omega$-strongly convex ($\omega>0$) if for any $  x, y \in {\mathbb{R}^n}$,
%\begin{equation*}
%  {(x - y)^T}(\nabla f(x) - \nabla f(y)) \ge \omega{\left\| {x - y} \right\|^2}.
%\end{equation*}
A function $f:{\mathbb{R}^n} \to \mathbb{R}$ is  $\omega$-strongly monotone ($\omega>0$) if for any $  x, y \in {\mathbb{R}^n}$
\begin{equation*}
  {(x - y)^T}( f(x) -  f(y)) \ge \omega{\left\| {x - y} \right\|^2}.
\end{equation*}
A function $f:{\mathbb{R}^n} \to \mathbb{R}^n$ is $\theta$-Lipschitz ($\theta>0$) if for any $  x, y \in {\mathbb{R}^n}$,
\begin{equation*}
  \left\| {f(x) - f(y)} \right\| \le \theta{\left\| {x - y} \right\|}.
\end{equation*}

\subsection{Problem Formulation}

Consider a multi-cluster game of $N$ clusters over an undirected graph $\mathcal{G}_0$. Cluster $j\in \{1,\dots, N\}$ consists of  $n_j$ players over an undirected graph $\mathcal{G}_j$, where $\mathcal{G}_j$ is a subgraph of $\mathcal{G}_0$.
Player $i$ in cluster $j$ has a continuously differentiable cost function  $f_i^j ( x^j_i,\textbf{x}^{-j}): \mathbb{R}^{q+\sum_{l=1, l \neq j }^N q n_l} \rightarrow \mathbb{R}$,
where $x_i^j \in \mathbb{R}^{q}$ is the decision of player $i$ in cluster $j$,
%$\bar{q}=\sum_{j=1}^N qn_j$,
$\textbf{x}^{-j}=col (\textbf{x}^1 ,\textbf{x}^2, \dots, \textbf{x}^{j-1}, \textbf{x}^{j+1}, \dots, \textbf{x}^N ) \in \mathbb{R}^{\sum_{l=1, l \neq j }^N q n_l}$,
$\textbf{x}^j=col( x^j_1 , \dots,  x^j_{n_j} ) \in \mathbb{R}^{q n_j} $.
Players in the same cluster are required to reach a common strategy.
%that is, $x_i^j = x_k^j, \, \forall i,k \in \{ 1, \dots, n_j \}$.
The objective of players in cluster $j$ is to minimize $ f^j \left(\textbf{x}^j,\textbf{x}^{-j} \right)$ by competing with other clusters, where $ f^j (\textbf{x}^j,\textbf{x}^{-j} )=\sum_{i=1}^{n_j} f_i^j ( x_i^j,\textbf{x}^{-j} )$ is the cost function of cluster $j$. Specifically, cluster $j$ faces the following multi-cluster game problem:
\begin{equation} \label{eq multi-cluster game}
\begin{split}
&\min_{\textbf{x}^j \in \mathbb{R}^{qn_j}} f^j \left(\textbf{x}^j,\textbf{x}^{-j} \right) \\
&s.t. \, x_i^j = x_k^j, \, \forall i,k \in \{ 1, \dots, n_j \}.
\end{split}
%where $ f^j (x^j,\textbf{x}^{-j} )=\sum_{i=1}^{n_j} f_i^j ( x^j,\textbf{x}^{-j} )$ is the  cost function of cluster $j$.
\end{equation}

%\eqref{eq multi-cluster game} is equivalent to the following multi-cluster game with equality constraints
%\begin{subequations} \label{eq multi-cluster game constraint}
%\begin{align}
%&\min_{\textbf{x}^j \in \mathbb{R}^{qn_j}} f^j \left(\textbf{x}^j,\textbf{x}^{-j} \right) \\
%&s.t.\,x_i^j = x_k^j, \, \forall i,k \in \{ 1, \dots, n_j \} \label{eq multi-cluster game constraint b}
%\end{align}
%\end{subequations}
%where
%%$x_i^j \in \mathbb{R}^q$ is the decision of player $i$ in cluster $j$, $\textbf{x}^j=col( x^j_1 , \dots,  x^j_{n_j} ) \in \mathbb{R}^{q n_j} $,
%$ f^j (\textbf{x}^j,\textbf{x}^{-j} )=\sum_{i=1}^{n_j} f_i^j ( x_i^j,x^{-j} )$.

The Nash equilibrium of the multi-cluster game \eqref{eq multi-cluster game} is defined as follows (see \cite{holt2004nash,deng2018distributedgeneralized,facchinei2010generalized}).
\begin{definition} \label{def GNE}
A strategy profile $\textbf{x}^*=(\textbf{x}^{j*},\textbf{x}^{-j*})$ is said to be a Nash equilibrium of the multi-cluster game \eqref{eq multi-cluster game} if  for all $j \in \{1,\dots, N\}$, we have
\begin{equation*}
f^j(\textbf{x}^{j*}, \textbf{x}^{-j*}) \leq f^j(\textbf{x}^{j}, \textbf{x}^{-j*}), ~\forall \, \textbf{x}^j: (\textbf{x}^{j}, \textbf{x}^{-j*}) \in \Omega
\end{equation*}
where $\Omega=\Omega^1\cap \dots \cap \Omega^N$ with $\Omega^j = \big\{\textbf{x}^j\in \mathbb{R}^{q n_j} : x_i^j = x_k^j, \, \forall i,k \in \{ 1, \dots, n_j \}\big\}$
%for all $j \in \{1,\dots, N\}$.
%\begin{equation*}
%f^j(\textbf{x}^{j*}, \textbf{x}^{-j*}) \leq f^j(\textbf{x}^{j}, \textbf{x}^{-j*}), ~\forall \, \textbf{x}^j \in \mathbb{R}^{qn_j}, \forall j \in \{1,\dots, N\}.
%\end{equation*}
%where $C=C^1 \times \cdots \times C^N\in \mathbb{R}^{\bar{q}}$, $C^j=C_1^1 \times \cdots \times C_{n_j}^N \in \mathbb{R}^{qn_j}$.
\end{definition}
Based on Definition \ref{def GNE}, a Nash equilibrium is a strategy profile on which cluster $j$ cannot reduce its cost $f^j(\textbf{x}^{j}, \textbf{x}^{-j})$ by unilaterally changing its own decision.

%A Nash equilibrium is said to be a Nash equilibrium if it is a solution of the variational inequality $VI\left( \mathbb{R}^{qn_j},F \right)$ (see \cite{facchinei201012}), where $F\left( \cdot \right) : \mathbb{R}^{\bar{q}} \rightarrow \mathbb{R}^{\bar{q}}$ is defined as
%\begin{align}
%F\left( \textbf{x} \right)&=col(\nabla_{{x}_1^1} f_1^1 ( x_1^1,\textbf{x}^{-1} ), \dots,
%\nabla_{{x}_{n_1}^1} f_{n_1}^1 ( x_{n_1}^1,\textbf{x}^{-1} ), \dots, \nonumber\\
%&\nabla_{{x}_1^N} f_1^N ( x_1^N,\textbf{x}^{-N} ), \dots,
%\nabla_{{x}_{n_N}^N} f_{n_N}^N ( x_{n_N}^N,\textbf{x}^{-N} ))
%\end{align}
%where $\textbf{x}=col(\textbf{x}^1, \dots, \textbf{x}^N) \in \mathbb{R}^{\bar{q}}$, $\bar{q}=\sum_{j=1}^N qn_j$.

%Our task is to design a distributed Nash equilibrium seeking algorithm for the high-order player \eqref{eq sys} such that the decisions of all players converge to the Nash equilibrium of the multi-cluster game \eqref{eq multi-cluster game}.

Some standard assumptions are given as follows.
\begin{assumption}\label{Ass digraphs}
Undirected graphs $\mathcal{G}_0, \dots, \mathcal{G}_N$ are connected.
\end{assumption}
\begin{assumption}\label{Ass cost fc convex}
The cost function $f_i^j ( x_i^j,\textbf{x}^{-j} )$ is convex in $x_i^j$, and the map $F(\textbf{x})$ is $\omega$-strongly monotone and $\theta$-Lipschitz in $\textbf{x}$,  where $F\left( \cdot \right) : \mathbb{R}^{\bar{q}} \rightarrow \mathbb{R}^{\bar{q}}$ is defined as
\begin{align}
F\left( \textbf{x} \right)= & col(\nabla_{{x}_1^1} f_1^1 ( x_1^1,\textbf{x}^{-1} ), \dots,
\nabla_{{x}_{n_1}^1} f_{n_1}^1 ( x_{n_1}^1,\textbf{x}^{-1} ), \dots, \nonumber\\
&\nabla_{{x}_1^N} f_1^N ( x_1^N,\textbf{x}^{-N} ), \dots,
\nabla_{{x}_{n_N}^N} f_{n_N}^N ( x_{n_N}^N,\textbf{x}^{-N} ))
\end{align}
with $\textbf{x}=col(\textbf{x}^1, \dots, \textbf{x}^N) \in \mathbb{R}^{\bar{q}}$ and $\bar{q}=\sum_{j=1}^N qn_j$.
\end{assumption}
%By Assumption \ref{Ass cost fc convex}, the cost function $f_i^j ( x_i^j,\textbf{x}^{-j} )$ is convex, which indicates that the variational Nash equilibrium of \eqref{eq multi-cluster game} exists.
Under Assumption \ref{Ass cost fc convex}, we have the following lemma about the Nash equilibrium.

%The following lemma is about the variational GNE of the multi-cluster game \eqref{eq multi-cluster game}.

\begin{lemma}\label{lemma.ne}
Suppose Assumption \ref{Ass cost fc convex} holds. $\textbf{x}=(\textbf{x}^{j},\textbf{x}^{-j})$ is a Nash equilibrium of the multi-cluster game \eqref{eq multi-cluster game} if and only if
%\begin{equation}\label{eq lemma.variational NE}
% \nabla_{\textbf{x}^{j}} f^j ( \textbf{x}^{j*},\textbf{x}^{-j*} )=0_{q}.
%\end{equation}
\begin{equation}\label{eq lemma.variational NE}
\begin{split}
&\sum_{i=1}^{n_j} \nabla_{x_i^j} f_i^j ( x_i^j,\textbf{x}^{-j} )=0_{q}\\
&\,x_i^j = x_k^j, \, \forall i,k \in \{ 1, \cdots, n_j \}.
\end{split}
\end{equation}

%\begin{subequations}\label{eq lemma.variational NE}
%\begin{align}
%& \nabla_{\textbf{x}^{j}} f^j ( \textbf{x}^{j*},\textbf{x}^{-j*} )=0_{qn_j} \\
%&\,x_i^j = x_k^j, \, \forall i,k \in \{ 1, \cdots, n_j \}, \forall j\in \{1,\dots, N\}. \label{eq lemma.variational NE b}
%\end{align}
%\end{subequations}
\end{lemma}
\emph{Proof:} Based on \cite[Theorem 3.9 and Theorem 4.8]{facchinei2010generalized}, the solution of the variational inequality $VI\left(\mathbb{R}^{\bar{q}},F \right)$ satisfis \eqref{eq lemma.variational NE}, and coincides with the Nash equilibrium of the multi-cluster game \eqref{eq multi-cluster game}.
\hfill $\Box$

Player $i$ in cluster $j$ has the following $n$th-order dynamics.
\begin{eqnarray}\label{eq sys}
x_i^{j(n)}=u_i^j %~ i \in \{1,...,N\},
\end{eqnarray}
where
$x_i^{j(n)}$ is the $n$th order derivative of $x_i^j$ with $n \geq 1$, and
$u_i^j\in \mathbb{R}^q$ is the control input.

The objective of this paper is to design a distributed algorithm for the high-order player \eqref{eq sys} such that the outputs of all players converge to the Nash equilibrium of the multi-cluster game \eqref{eq multi-cluster game}, which means the high-order player \eqref{eq sys} can carry out the multi-cluster game task \eqref{eq multi-cluster game} autonomously.

%\begin{remark}
%In contrast to well-studied RAPs involving first-order dynamics (e.g. \cite{Xiao06,Lakshmanan08,Wei13,Cherukuri15}) or second-order dynamics (e.g. \cite{DengLY18,Deng}), the high-order dynamics is taken into account for every agent in this formulation.
%Besides, different from the distributed optimization problems of high-order systems considered in \cite{DengZH16,Zhang15}, this formulation contains  network resource constraints.
%Without considering the high-order dynamics of agents and/or network resource constraints, existing distributed  algorithms, including \cite{Xiao06,Lakshmanan08,Beck14,Wei13,Cherukuri15,DengLH17,DengLY18,Deng,DengZH16,Zhang15}, cannot guarantee that the high-order agent \eqref{sys} converges to the optimal allocation of RAP \eqref{pro}.
%Additionally, the cost functions and their gradients for optimization are basically nonlinear, which together with the high-order dynamics of agents  imply that it is not easy to  design distributed algorithms and analyze their convergence.
%%Each agent only knows the local resource information rather than the global one, different from those in \cite{Lakshmanan08,Xiao06,Wei13,Johari05,Cherukuri15}.
%%          as practical cases like power grid \cite{Zhang12}. The network resource is , d, and reveals some . Additionally, the network resource is necessary to be known by all agents or at least one agent in many results like [16], [15], [19], [23], but .
%\end{remark}

%\begin{assumption}\label{ass.g}
%The digraph $\mathcal{G}$ is strongly connected and weight-balanced.
%\end{assumption}

\begin{remark}
Our formulation can be viewed as extensions of  distributed optimization problems and noncooperative game problems: when $N=1$, the problem \eqref{eq multi-cluster game} is reduced to the distributed optimization problems studied in \cite{liang2019exponential,yang2017distributed,liy2019,he2019continuous,kia2015distributed,Zhang17}; %\cite{liang2018distributed}, \cite{lakshmanan2008decentralized}, \cite{bretthauer1997quadratic}, \cite{Xiao06};
when $n_j=1,\, \forall j \in \{1,\dots, N\}$, the problem \eqref{eq multi-cluster game} is degraded into the noncooperative games of $N$ players investigated in \cite{ZhangLWJ19,deng2021distributedEL,bianchiG2021,RomanoP20},
%\cite{ deng2018distributed}, \cite{deng2019distributedEL}, \cite{D.Gadjov2019noncoop-game}, \cite{C.De2020noncoop-game}.
Therefore, the multi-cluster game \eqref{eq multi-cluster game} involves cooperative and competitive behaviors of the players simultaneously: players in the same cluster collectively optimize the cost function of the cluster, while players in different clusters selfishly minimize their own cost functions of the clusters that they belong to.
Moreover, without involving the high-order dynamics,
%which enables the high-order player \eqref{eq sys} to complete multi-cluster game \eqref{eq multi-cluster game} autonomously.
%High-order dynamics of players lead to the fact that all existing Nash equilibrium seeking algorithms for multi-cluster games (see \cite{YE2018266}, \cite{Ye2019AUnifiedStrategy}, \cite{meng2020linear}, \cite{pang2020gradient},
 existing Nash equilibrium seeking algorithms for multi-cluster games (such as \cite{YE2018266,ZENG201920,Ye2019AUnifiedStrategy,meng2020linear,pang2020gradient}) cannot control the high-order player \eqref{eq sys}  to  accomplish  multi-cluster game task \eqref{eq multi-cluster game} autonomously.
 Also, the high-order dynamics of players and the nonlinearity of cost functions make it difficult to design and analyze distributed game algorithms.
%collaboratively where players in the same cluster are involved in a resource allocation problem and clusters act as players in a noncooperative game. In the multi-cluster game \eqref{eq multi-cluster game},
%\cite{deng2020resourceHighOrder}, \cite{Xiao06}, \cite{L.Jin2018resource}, \cite{R.Shang2016resourcr}, \cite{M.Zargham2014resource}.
%Therefore, the multi-cluster game considered in this paper involves the noncooperative game and resource allocation problem as special cases.
\end{remark}

%\begin{remark}
%All existing Nash equilibrium seeking algorithms for multi-cluster games involve  players (see \cite{YE2018266}, \cite{Ye2019AUnifiedStrategy}, \cite{meng2020linear}, \cite{pang2020gradient}, \cite{ZENG201920}), which cannot guarantee the high-order player \eqref{eq sys} converges to the Nash equilibrium of multi-cluster game \eqref{eq multi-cluster game}. In this formulation, the high-order dynamics is taken into account for every player, which enables the high-order player \eqref{eq sys} to complete multi-cluster game \eqref{eq multi-cluster game} autonomously.
%\end{remark}

\section{Main Results}\label{sec.r}

In this section, we propose a distributed Nash equilibrium seeking algorithm for the multi-cluster game \eqref{eq multi-cluster game} with high-order player \eqref{eq sys} in Subsection \ref{ssec.algorithm}, and then analyze its convergence in Subsection \ref{ssec.convergence}.

\subsection{Distributed Algorithm Design}\label{ssec.algorithm}

This subsection provides a distributed Nash equilibrium seeking algorithm for the multi-cluster game \eqref{eq multi-cluster game} with high-order player \eqref{eq sys}.

Before giving our algorithm, the following characteristic polynomial associated with real coefficients $(k_1,\ldots,k_{n-1})$ is defined such that its roots are in the open left half plane (LHP).
\begin{align}\label{eq characteristic polynomial}
p(s) :=s^{n-1} +k_{n-1} s^{n-2} +\ldots +k_{2} s +k_{1}
\end{align}
%where $k_1 >0$.
which implies that the following companion matrix $A$ is Hurwitz.
\begin{align}\label{eqA}
A=\left[
  \begin{array}{ccc}
    0_{n-2} & \vline & I_{n-2}  \\
\hline
-k_1 & \vline &
 \left[
   \begin{array}{ccc}
     -k_2 &  \ldots & -k_{n-1}
   \end{array}
 \right]
  \end{array}
\right].
\end{align}

The following lemma is about the companion matrix $A$, which is used later (see \cite[Theorem 5.6]{Chen99}). %we have the following result, based on \cite[Theorem 4.6]{Khalil02}.
\begin{lemma}\label{lemma.cp}
There is a positive definite symmetric matrix $P_1$ such that $P_1A +A^T P_1 =-I_{n-1}$ is satisfied, where $P_1:=[p_{ij}]_{(n-1) \times (n-1)}$.
\end{lemma}

%Consider the high-order player \eqref{eq sys},
The distributed Nash equilibrium seeking algorithm for player $i$ in cluster $j$ is designed as follows.
\begin{subequations}\label{eq algorithm}
\begin{align}
%&&\left\{\begin{split}
u_i^j =& -\sum_{l=1}^{n-1} \varepsilon^{n-l} k_l x_i^{j(l)} -y_i^j - \nabla_{x_i^j} f_i^j(x_i^j,\hat{\textbf{x}}^{-j})  \label{alga}\\
\dot {y}_i^j =& \kappa_1 \sum_{k=1}^{n_j} a_{ik}^j (x_i^j - x_k^j), \quad y_i^j(0)=0_q  \\
\dot{\hat{x}}_{mn}^{ji} =& -\kappa_2 \big(\sum_{u=1}^{N} \sum_{v=1}^{n_u} \bar{a}_{iv}^0 (\hat{x}_{mn}^{ji}-\hat{x}_{mn}^{uv}) + \bar{a}_{in}^0 (\hat{x}_{mn}^{ji}-x_n^m) \big)   \label{algc}
%\end{split}\right.
\end{align}
\end{subequations}
where
$\hat{x}_{mn}^{ji} \in \mathbb{R}^{q}$ %$=[ ( \hat{x}_{mn,1}^{ji} )^T, \dots, ( \hat{x}_{mn,q}^{ji} )^T ]^T \in \mathbb{R}^{q}$
is the estimation of player $i$ in cluster $j$ on $x_n^m$ of player $n$ in cluster $m$ with $ m\in \{ 1, 2, \dots, N \}$ and $n \in  \{ 1,\dots, n_m \}$,  $\hat{\textbf{x}}^{-j}=col(\hat{x}_{11}^{ji}, \dots, \hat{x}_{1n_1}^{ji}, \dots, \hat{x}_{(j-1) 1 }^{ji}, \dots, \hat{x}_{(j-1) n_{(j-1)} }^{ji},$
$\hat{x}_{(j+1)1 }^{ji}, \dots, \hat{x}_{(j+1) n_{(j+1)} }^{ji},\dots, \hat{x}_{N1}^{ji}, \dots, \hat{x}_{N n_N }^{ji} )$,  $\bar{a}_{iv}^0=a_{\left(\sum_{p=0}^{j-1}n_p+i\right)\left(\sum_{q=0}^{u-1}n_q+v\right)}^0$,  $\bar{a}_{in}^0=a_{\left(\sum_{p=0}^{j-1}n_p+i\right)\left(\sum_{q=0}^{m-1}n_q+n\right)}^0$ with $a_{(\cdot)(\cdot)}^0$ being the element of the adjacency matrix $\mathcal{A}_0$ of $\mathcal{G}_0$,
$a_{ik}^j$ is the element of the adjacency matrix $\mathcal{A}_j$ of $\mathcal{G}_j$,
$k_1,\ldots, k_{n-1}$ are the coefficients of the characteristic polynomial \eqref{eq characteristic polynomial} with roots in the open LHP,
$\varepsilon > \sqrt[n]{\frac{3}{4}(\theta \bar{a}_1 + \bar{a}_1)}$,
$\kappa_2 > \frac{12\omega \varepsilon^{n}\lambda_{max}^2(P_2)+\theta^2 k_1 + 2 \omega\mu^2 \theta^2 + \omega \theta (n-1)}{2\omega \varepsilon^{n-1} \lambda_{min}(Q)}$, $\mu > \frac{k_1}{\omega} + \frac{2n-1}{4}$,
%$P_2$ and $Q$ are symmetric positive-define matrices defined in Subsection \ref{ssec.convergence},
$P_2$ and $Q$ are symmetric positive-define matrices satisfying $P_2 (\bm{\mathcal{L}} \otimes I_{\bar{N}}+ M) + (\bm{\mathcal{L}} \otimes I_{\bar{N}}+ M)^T P_2 =Q$,
$M=diag\{\bar{a}^0_{in}\}$,
$\bm{\mathcal{L}}$ is the Laplacian matrix of $\mathcal{G}_0$,
\begin{align*}
&\bar{a}_1 = max\{(2p_{1(n-1)}+k_2)^2,\cdots,(2p_{(n-2)(n-1)}+k_{n-1})^2,\\
&\qquad \qquad \quad (2p_{(n-1)(n-1)}+1)^2\}, \\
&0<\kappa_1<min\{\frac{\omega k_1 }{2 \varepsilon^{n-2}(3n+ \varepsilon \mu^2  + 2\varepsilon \mu k_1 \| \textbf{L}\| -3)}, \\
&\qquad \qquad \qquad ~\frac{1}{\|\textbf{L}\|^2 \mu^2 k_{max}^2},
 \frac{4\omega \mu + 2\omega n +4k_1-\omega}{2\omega \|\textbf{L}\|^2 \mu^2\varepsilon^{n-1}}
\},
\end{align*}
$p_{i(n-1)}$ is the last element of the $i$th row vector of matrix $P_1$ defined in Lemma \ref{lemma.cp},
$k_{max}=max\{1,k_2,\ldots, k_{n-1}\}$, $\textbf{L}=diag \{L^1, \dots, L^N \}$ with $L^j$ being the Laplacian matrix of $\mathcal{G}_j$.

\begin{remark}
The algorithm \eqref{eq algorithm} is made up of four parts:\\
(i) $x_i^{j(l)}$ as the state feedback for stabilizing the high-order player \eqref{eq sys};\\
(ii)  $\nabla_{x_i^j} f_i^j ( x_i^j, \hat{\textbf{x}}^{-j} )$ for seeking the Nash equilibrium;\\
%(iii) $y_i^j$ for the satisfaction of \eqref{eq multi-cluster game constraint b};
(iii) $y_i^j$ for the consistency of the decisions of players in cluster $j$;\\
%(iv), a auxiliary variable, for information sharing among players;
and (iv) ${\hat{x}} _{mn}^{ji}$ for estimating $x_n^m$.
%and $\mu_i$ for the satisfaction of the optimal condition \eqref{optb}.
%By the algorithm \eqref{alg}, agent $i$ transmits $\lambda_i$ and $\mu_i$ to its neighbors.
%Besides, the agents do not share their cost functions, subgradients and decisions with their neighbors, which implies that the algorithm \eqref{alg} privatizes these information.
\end{remark}

\begin{remark}
In contrast to many existing multi-cluster game algorithms that require every player to have access to the decisions of all players, such as \cite{YE2018266,ZENG201920,Ye2019AUnifiedStrategy,pang2020gradient}, players with the algorithm \eqref{eq algorithm} only exchange necessary information with their neighbors.
%in cluster $j$, but does not share its decision with players in other clusters.
Besides, the cost functions and gradients of players are not shared with any other players, which signifies that the algorithm \eqref{eq algorithm} is conducive to protect these information.
%all players only exchange with their neighbors.
\end{remark}

\subsection{Convergence Analysis}\label{ssec.convergence}

The convergence of algorithm \eqref{eq algorithm} is analyzed in this subsection.

Let
\begin{align*}
\textbf{x}=& col(\textbf{x}^1,\ldots,\textbf{x}^N) \in \mathbb{R}^{\bar{q}}\\
\textbf{x}^j=& col(x_i^j,\ldots,x_{n_j}^j) \in \mathbb{R}^{qn_j}\\
\textbf{y}=& col(\textbf{y}^1,\ldots,\textbf{y}^N) \in \mathbb{R}^{\bar{q}}\\
\textbf{y}^j=& col(y_i^j,\ldots,y_{n_j}^j) \in \mathbb{R}^{qn_j}\\
\textbf{x}^{(l)}=& col(\textbf{x}^{1(l)},\ldots, \textbf{x}^{N(l)}) \in \mathbb{R}^{\bar{q}} \\
\textbf{x}^{j(l)}=& col(x_1^{j(l)},\ldots, x_{n_j}^{j(l)}) \in \mathbb{R}^{qn_j}\\
\hat{\textbf{x}} =&  col (\hat{\textbf{x}}^{11} , \dots, \hat{\textbf{x}}^{Nn_N} ) \in \mathbb{R} ^{\bar{N} \bar{q}} \\
\hat{\textbf{x}}^{ji}=& col ( \hat{x}_{11}^{ji} , \dots, \hat{x}_{N_{n_N}}^{ji}) \in \mathbb{R} ^{\bar{q}}
\end{align*}
where $l \in \{1, \ldots, n\}$ and $\bar{N} =  \sum_{j=1}^{N} n_j$.

Combining \eqref{eq sys} with \eqref{eq algorithm}, we have
\begin{subequations}\label{eq equilibrium point}
\begin{align}
\dot {\textbf{x}}\! =& \textbf{x}^{(1)}  \\
\vdots & \nonumber \\
\textbf{x}^{(n)}\! =& -\sum_{l=1}^{n-1} \varepsilon^{n-l} k_l \textbf{x}^{(l)} -\textbf{y} - F(\hat{\textbf{x}}) \hfill \\
\dot {\textbf{y}}\! =& \kappa_1((\textbf{L} \otimes I_q)\textbf{x}), \quad \textbf{y}(0)=0_{\bar{q}}   \\
\dot{\hat{\textbf{x}}} \!=& \!- \!\kappa_2(((\bm{\mathcal{L}} \otimes I_{\bar{N}}) \otimes I_q)\hat{\textbf{x}}\! +\! (M \!\otimes \! I_q) (\hat{\textbf{x}}\! - \!1_{\bar{N}} \otimes \textbf{x}))
\end{align}
\end{subequations}
where
%$M=diag\{\bar{a}_{in}\}$,
%$\bm{\mathcal{L}}$ is the Laplacian matrix of $\mathcal{G}_0$,
$F\left( \hat{\textbf{x}} \right)=col(\nabla_{{x}_1^1} f_1^1 ( x_1^1,\hat{\textbf{x}}^{-1}),
 \dots, \nabla_{{x}_{n_1}^1} f_{n_1}^1 ( x_{n_1}^1,\hat{\textbf{x}}^{-1} ), $
$ \dots, \nabla_{{x}_1^N} f_1^N ( x_1^N,\hat{\textbf{x}}^{-N} ), \dots,
\nabla_{{x}_{n_N}^N} f_{n_N}^N ( x_{n_N}^N,\hat{\textbf{x}}^{-N} ))$.

The following lemma is about $F\left( \hat{\textbf{x}} \right)$, which is used later.
\begin{lemma}\label{lem.F}
Under Assumption \ref{Ass cost fc convex}, $F\left( \hat{\textbf{x}} \right)$ is $\theta$-Lipschitz.
\end{lemma}
\emph{Proof:} Based on the previous definitions, we have
$F( \hat{\textbf{x}}^{ji} )=
col(\nabla_{{x}_1^1} f_1^1 ( x_1^1,\hat{\textbf{x}}^{-j} ), \dots,
\nabla_{{x}_{n_1}^1} f_{n_1}^1 ( x_{n_1}^1,\hat{\textbf{x}}^{-j} ), \dots, $
$
\nabla_{{x}_1^N} f_1^N ( x_1^N,\hat{\textbf{x}}^{-j} ), \dots,
\nabla_{{x}_{n_N}^N} f_{n_N}^N ( x_{n_N}^N,\hat{\textbf{x}}^{-j} ))$.
Define $\hat{\textbf{y}}^{-j}$, $\hat{\textbf{y}}^{ji}$ and $\hat{\textbf{y}}$ in the same way as $\hat{\textbf{x}}^{-j}$, $\hat{\textbf{x}}^{ji}$ and $\hat{\textbf{x}}$.
Then, by Assumption \ref{Ass cost fc convex}, it is obvious that $\| \nabla_{{x}_i^j} f_i^j ( x_i^j,\hat{\textbf{x}}^{-j} ) - \nabla_{{x}_i^j} f_i^j ( x_i^j,\hat{\textbf{y}}^{-j} )\| \leq
\|F (\hat{\textbf{x}}^{ji} )-F (\hat{\textbf{y}}^{ji} )\| \leq
\theta \| \hat{\textbf{x}}^{ji}-\hat{\textbf{y}}^{ji} \|$, $\forall j\in \{1,\dots, N\},\,\forall i,k\in \{1,\dots, n_j\}$,
%$\hat{\textbf{y}}^{-j}=col(\hat{y}_{11}^{ji}, \dots, \hat{y}_{1n_1}^{ji}, \dots, \hat{y}_{(j-1) 1 }^{ji}, \dots, \hat{y}_{(j-1) n_{(j-1)} }^{ji}$,
%$\hat{\textbf{x}}^{ji}= col ( \hat{x}_{11}^{ji} , \dots, \hat{x}_{N_{n_N}}^{ji} )$,
Accordingly, the following inequalities are obtained: $ \|F (\hat{\textbf{x}})-F (\hat{\textbf{y}}) \|^2 =
\sum_{j=1}^{N} \sum_{i=1}^{n_j}\| \nabla_{{x}_i^j} f_i^j ( x_i^j,\hat{\textbf{x}}^{-j} ) - \nabla_{{x}_i^j} f_i^j ( x_i^j,\hat{\textbf{y}}^{-j} )\|^2 \leq
\sum_{j=1}^{N} \sum_{i=1}^{n_j} \theta^2 \| \hat{\textbf{x}}^{ji}-\hat{\textbf{y}}^{ji} \|^2 \leq
\theta^2 \|\hat{\textbf{x}}-\hat{\textbf{y}} \|^2 $, which implies that $F\left( \hat{\textbf{x}} \right)$ is $\theta$-Lipschitz.
\hfill $\Box$

Next, the relationship between the equilibrium point of \eqref{eq equilibrium point} and the Nash equilibrium of the multi-cluster game \eqref{eq multi-cluster game} is analyzed, which yields the following result.
\begin{theorem}\label{theorem.eq point}
%Under Assumptions \ref{Ass digraphs} and \ref{Ass cost fc convex},
%if $(\textbf{x}^*,\textbf{y}^*,\hat{\textbf{x}}^*)$ is an equilibrium point of \eqref{eq equilibrium point}, $\textbf{x}^*$ is a Nash equilibrium of the multi-cluster game \eqref{eq multi-cluster game}.
%Moreover, if $\textbf{x}^*$ is a equilibrium point of the multi-cluster game \eqref{eq multi-cluster game}, there exist $\textbf{y}^* \in \mathbb{R}^{\bar{q}}$ and $\hat{\textbf{x}}^* \in \mathbb{R}^{\bar{q}}$ such that $(\textbf{x}^*,\textbf{y}^*,\hat{\textbf{x}}^*)$ is an equilibrium point of \eqref{eq equilibrium point}.
Under Assumptions \ref{Ass digraphs} and \ref{Ass cost fc convex},
$\textbf{x}^*$ is a Nash equilibrium of the multi-cluster game \eqref{eq multi-cluster game} if and only if there exist $\textbf{y}^* \in \mathbb{R}^{\bar{q}}$ and $\hat{\textbf{x}}^* \in \mathbb{R}^{\bar{q}}$ such that $(\textbf{x}^*,\textbf{y}^*,\hat{\textbf{x}}^*)$ is an equilibrium point of \eqref{eq equilibrium point}.
\end{theorem}
\emph{proof:}
(i)
The equilibrium point of \eqref{eq equilibrium point} satisfies the following equations:
\begin{subequations}\label{eq equilibrium point 0}
\begin{align}
%&&\left\{\begin{split}
 \textbf{x}^{*(1)}=&0_{\bar{q}}  \label{eq equilibrium point 0 a} \\
\vdots & \nonumber\\
%\dot{x}^{(n-1)} =& {x}^{(n)},\\
 -\sum_{l=1}^{n-1} \varepsilon^{n-l} k_l \textbf{x}^{*(l)} -\textbf{y}^* - F(\hat{\textbf{x}}^*)=&0_{\bar{q}} \label{eq equilibrium point 0 b} \\
 (\textbf{L} \otimes I_q)\textbf{x}^*=&0_{\bar{q}} \label{eq equilibrium point 0 c} \\
-(((\bm{\mathcal{L}} \! \otimes \! I_{\bar{N}}) \otimes  I_q)\hat{\textbf{x}}^* \!+\! (M \! \otimes \!  I_q)(\hat{\textbf{x}}^* \! - \! 1_{\bar{N}}  \otimes  \textbf{x}^*) ) \! =&0_{\bar{N} \bar{q}}.\label{eq equilibrium point 0 d}
%\end{split}\right.
\end{align}
\end{subequations}
%$\hat{\textbf{x}}^* = 1_{\bar{N}}\otimes\textbf{x}^* $

%Due to $\textbf{y}(0)=0_{\bar{q}}$, it follows from \eqref{eq equilibrium point c} that $(1_{n_j}^T \otimes I_q)\textbf{y}^j=0_q$.
It results from \eqref{eq equilibrium point 0 d} that
$ - ((\bm{\mathcal{L}} \otimes  I_{\bar{N}}) \otimes I_q + (M \otimes  I_q))\hat{\textbf{x}}^* = - ((\bm{\mathcal{L}} \otimes  I_{\bar{N}}) \otimes I_q +(M \otimes  I_q))(1_{\bar{N}} \otimes \textbf{x}^*) $.
Then, because $\mathcal{G}^0$ is a connected undirected graph and $M$ is a diagonal matrix with at least one diagonal element being positive, $(\bm{\mathcal{L}} \otimes  I_{\bar{N}}) \otimes I_q + (M \otimes  I_q)$ is positive define, which implies that
%$\hat{x}_{mn}^{ji*}=x_n^{m*}$,
$\hat{\textbf{x}}^*=1_{\bar{N}} \otimes \textbf{x}^*$,
i.e., $F(\hat{\textbf{x}}^*)=F(1_{\bar{N}} \otimes \textbf{x}^*)$.
Besides, since $\mathcal{G}^1, \dots, \mathcal{G}^N$ are undirected and connected graphs,
i.e.,
$\textbf{L} 1_{\bar{N}}=0_{\bar{N}}$,
$1_{\bar{N}}^T \textbf{L}=0_{\bar{N}}^T $,
%i.e., $L^j1_{n_j}=0_{qn_j}$,
%$1_{n_j}^T L^j=0_{qn_j}^T,\,j \in \{1,\dots, N\}$,
%$\textbf{L} 1_{\bar{N}}=0_{\bar{N}}$,
%$1_{\bar{N}}^T \textbf{L}=0_{\bar{N}}^T $, $\bm{\mathcal{L}} 1_{\bar{N}}=0_{\bar{N}}$, and
%$1_{\bar{N}}^T \bm{\mathcal{L}}=0_{\bar{N}}^T $,
\eqref{eq equilibrium point 0 c} yields that $x_i^{j*} = x_k^{j*}, \, \forall i,k \in \{ 1, \cdots, n_j \}$,  and $(1_{n_j}^T \otimes I_q)\textbf{y}^{j}=0_q$, $\forall j\in \{1,\dots, N\}$, by reason of $\textbf{y}(0)=0_{\bar{q}}$.
In addition, it follows from \eqref{eq equilibrium point 0 a} and \eqref{eq equilibrium point 0 b} that
$(1_{n_j}^T \otimes I_q)\textbf{y}^{j*} + (1_{n_j}^T \otimes I_q) \sum_{i=1}^{n_j} \nabla_{{x}_i^j} f_i^j ( x_i^{j*},\textbf{x}^{-j*} ) =0_q,  \,\forall j\in \{1,\dots, N\} $.
Further, we have
$\sum_{i=1}^{n_j} \nabla_{{x}_i^j} f_i^j ( x_i^{j*},\textbf{x}^{-j*} ) =0_q$.
%$\nabla_{\textbf{x}^{j}} f^j ( \textbf{x}^{j*},\textbf{x}^{-j*} )=0_{qn_j}$.
Therefore, according to Lemma \ref{lemma.ne}, $\textbf{x}^*$ is a Nash equilibrium of the multi-cluster game \eqref{eq multi-cluster game}.

(ii)
Conversely, if $\textbf{x}^*$ is a Nash equilibrium of the multi-cluster game \eqref{eq multi-cluster game}, we have
$\sum_{i=1}^{n_j} \nabla_{{x}_i^j} f_i^j ( x_i^j,\textbf{x}^{-j} ) =0_q$,
%$\nabla_{\textbf{x}^{j}} f^j ( \textbf{x}^{j*},\textbf{x}^{-j*} )=0_{qn_j}$
and $x_i^{j*}=x_k^{j*},\,\forall j\in \{1,\dots, N\},\,\forall i,k\in \{1,\dots, n_j\}$.
Take $\hat{x}_{mn}^{ji*}=x_n^{m*}$ and $\textbf{y}^* = F(\hat{\textbf{x}}^*)$.
Thus \eqref{eq equilibrium point 0} holds.
\hfill $\Box$
%\end{proof}

Theorem \ref{theorem.eq point} shows that if \eqref{eq equilibrium point} converges to its equilibrium points,  high-order player \eqref{eq sys} approaches the Nash equilibrium of the multi-cluster game \eqref{eq multi-cluster game}.
Consequently, we can obtain the following result by analyzing the convergence of \eqref{eq equilibrium point}.

\begin{theorem}\label{the.1}
Under Assumptions \ref{Ass digraphs} and \ref{Ass cost fc convex}, the high-order player \eqref{eq sys} with the algorithm \eqref{eq algorithm} globally exponentially converges to the Nash equilibrium of the multi-cluster game \eqref{eq multi-cluster game}.
\end{theorem}

\emph{Proof:}
We complete the proof in two steps.

\emph{Step 1: Coordinate transformations for \eqref{eq equilibrium point}.}

Make the following coordinate transformation.
\begin{align*}
\bar{\textbf{x}}=& \textbf{x}-\textbf{x}^*\\
\bar{\textbf{y}}=& \textbf{y}-\textbf{y}^*\\
\bar{\textbf{x}}^{(l)}=& \textbf{x}^{(l)}-\textbf{x}^{*(l)}\\
\bar{\hat{\textbf{x}}}=& \hat{\textbf{x}}-1_{\bar{N}}\otimes\textbf{x}
\end{align*}
where $l \in \{1, \ldots, n-1\}$.

\eqref{eq equilibrium point} and \eqref{eq equilibrium point 0} yield that
\begin{subequations}\label{eq transform}
\begin{align}
%&&\left\{\begin{split}
\dot {\bar{\textbf{x}}} =& \bar{\textbf{x}}^{(1)}\\
\vdots & \nonumber\\
%\dot{x}^{(n-1)} =& {x}^{(n)},\\
\bar{\textbf{x}}^{(n)} =& -\sum_{l=1}^{n-1} \varepsilon^{n-l} k_l \bar{\textbf{x}}^{(l)} - \bar{\textbf{y}} - h \\
\dot {\bar{\textbf{y}}} =& \kappa_1((\textbf{L} \otimes I_q)\bar{\textbf{x}})  \\
\dot{\bar{\hat{\textbf{x}}}} =& - \kappa_2((\bm{\mathcal{L}} \otimes I_{\bar{N}}) \otimes I_q +(M \otimes I_q))\bar{\hat{\textbf{x}}}
- 1_{\bar{N}}\otimes \dot{\bar{\textbf{x}}}
%\end{split}\right.
\end{align}
\end{subequations}
where $h=F(\hat{\textbf{x}})-F(\hat{\textbf{x}}^*)$.

With the above transformation, the equilibrium point of \eqref{eq transform} is the origin.

Let
\begin{align*}
\tilde{\textbf{x}} =& col(\tilde{\textbf{x}}^{(1)}, \ldots, \tilde{\textbf{x}}^{(n-1)}) \\
\tilde{\textbf{x}}^{(l)}=& col(\tilde{\textbf{x}}^{1(l)},\ldots, \tilde{\textbf{x}}^{N(l)}) \\
\tilde{\textbf{x}}^{j(l)}=& col(\tilde{x}_1^{j(l)},\ldots, \tilde{x}_{n_j}^{j(l)})
\end{align*}
where
$\tilde{x}_i^{j(l)} = \frac{1}{\varepsilon^{l}} {\bar x}_i^{j(l)}$ with
$l \in \{1, \ldots, n-1\}$.

Then \eqref{eq transform} can be rewritten as
\begin{subequations}\label{eq transform final}
\begin{align}
%&&\left\{\begin{split}
\dot {\bar{\textbf{x}}} =& \varepsilon \tilde{\textbf{x}}^{(1)}\\
%\dot{x}^{(n-1)} =& {x}^{(n)},\\
\dot{\tilde{\textbf{x}}} =& -\varepsilon(A \otimes I_{\bar{N}q})\tilde{\textbf{x}} -  \frac{1}{\varepsilon^{n-1}} (b\otimes I_{\bar{N}q}) (\bar{\textbf{y}} + h) \\
\dot {\bar{\textbf{y}}} =& \kappa_1((\textbf{L} \otimes I_q)\bar{\textbf{x}})  \\
\dot{\bar{\hat{\textbf{x}}}} =& - \kappa_2((\bm{\mathcal{L}} \otimes I_{\bar{N}}) \otimes I_q+(M \otimes I_q))\bar{\hat{\textbf{x}}}
- 1_{\bar{N}}\otimes \dot{\bar{\textbf{x}}}
\end{align}
\end{subequations}
where $b=\begin{bmatrix}
0_{n-2}^T & 1
\end{bmatrix}^T$ and $A$ is defined in \eqref{eqA}.

It is obvious that \eqref{eq transform} and \eqref{eq transform final} are equivalent.

Utilize the following orthogonal transformation,
\begin{subequations}
\label{eq orthogonal transformation}
\begin{align}
{\chi } =& col({\chi _{1}},{\chi _{2}}) = ({\begin{bmatrix}
r&R
\end{bmatrix}^T}  \otimes {I_q}){\bar {\textbf{x}}}\\
{\chi^{(l)} }=& col({\chi _{1}^{(l)} },{\chi _{2}^{(l)} }) = ({\begin{bmatrix}
r&R
\end{bmatrix}^T}\otimes {I_q}) {{\tilde {\textbf{x}}}^{(l)}}\\
\eta  =& col({\eta _1},{\eta _2}) = ({\begin{bmatrix}
r&R
\end{bmatrix}^T}\otimes {I_q}))\bar {\textbf{y}}
\end{align}
\end{subequations}
where
${\chi _{1}},   \chi _{1}^{(l)}, {\eta _1} \in \mathbb{R}^{q}$,
$ {\chi _{2}},  \chi _{2}^{(l)}, {\eta _2} \in \mathbb{R}^{(\bar{N}-1)q}$,
$l \in \{1, \ldots, n-1\}$,
$r = \frac{1}{{\sqrt {\bar{N}} }}{1_{\bar{N}}},{r^T}R = 0_{\bar{N}}^T,{R^T}R = {I_{\bar{N} - 1}}$ and $R{R^T} = {I_{\bar{N}}} - \frac{1}{\bar{N}}{1_{\bar{N}}}1_{\bar{N}}^T$.

Without loss of generality, let $q = 1$ for simplicity, and let
\begin{align*}
{\tilde{\chi} _{1}} =& col(\chi _{1}^{(1)},\ldots, \chi _{1}^{(n-1)})\\
{\tilde{\chi} _{2}} =&  col(\chi _{2}^{(1)},\ldots, \chi _{2}^{(n-1)}).
%{\tilde{\chi} _{y1}} =& col(\chi _{y1}^{(1)},\ldots, \chi _{y1}^{(n-1)}),\quad
%{\tilde{\chi} _{y2}} = col(\chi _{y2}^{(1)},\ldots, \chi _{y2}^{(n-1)}),
\end{align*}

Thus \eqref{eq transform final} can be described as
\begin{subequations}\label{eq orthogonal transformation final}
\begin{align}
%&&\left\{\begin{split}
{{\dot \chi }_{1}} =& \varepsilon \chi _{1}^{(1)}\\
{\dot {\tilde{\chi}} _{1}} =&    \varepsilon A \tilde{\chi} _{1} - \frac{1}{\varepsilon^{n-1}} b (\eta_1 +r^T h)\\
{{\dot \eta }_1} =& 0 \\
{{\dot \chi }_{2}} =& \varepsilon \chi _{2}^{(1)}\\
{\dot {\tilde{\chi}} _{2}} =&    \varepsilon (A \otimes I_{\bar{N}-1} ) \tilde{\chi} _{2} - \frac{1}{\varepsilon^{n-1}} (b \otimes I_{\bar{N}-1} ) (\eta_2 +R^T h)\\
{{\dot \eta }_2} =& \kappa_1 R^T \textbf{L} R  \chi _{2}\\
\dot{\bar{\hat{\textbf{x}}}} =& - \kappa_2((\bm{\mathcal{L}} \otimes I_{\bar{N}} ) + M )\bar{\hat{\textbf{x}}}
-1_{\bar{N}}\otimes \dot{\bar{\textbf{x}}}. \label{eq orthogonal transformation final g}
\end{align}
\end{subequations}

Obviously, $\textbf{x}$ approaches the  Nash equilibrium of
the multi-cluster game \eqref{eq multi-cluster game} if \eqref{eq orthogonal transformation final} tends to the origin. Consequently, our next task is to analyze the convergence of \eqref{eq orthogonal transformation final}.

%Let $c_1=\varepsilon^{n-1}$, $c_2=\frac{\omega}{2\theta}$, $c_7=\varepsilon^n$, $c_8 = \sqrt{\frac{1}{\varepsilon^{2n-2}}}$, $c_{10} = \frac{\omega}{4}$, $k_1=c_3=c_4=c_5=c_6=c_9=1$, $\varepsilon>0$, $\kappa_2>\frac{ \varepsilon^{n+1}}{\lambda_{min}(Q_1)}$, $c>\frac{2\theta^2 + \omega \theta \mu_2^2 + \omega \theta(n-1)}{2\varepsilon^{n-1} \omega (\kappa_2 \lambda_{min}(Q_1)-\varepsilon^{n+1})}$, $\mu_1=\sqrt{\frac{1}{\varepsilon^{2n}(n-1)}}$, $\mu_2>(\frac{\theta}{2\varepsilon^{2n}(n-1)}+\frac{n}{2}+\frac{2}{\omega})\sqrt{\varepsilon^{2n}(n-1)}$, $0<a<min\{\frac{2\omega \mu_2 \sqrt{\varepsilon^{2n}(n-1)}-\omega \theta - \omega \varepsilon^{2n}(n-1)^2 - \omega \varepsilon^{2n}(n-1) - 4\varepsilon^{2n}(n-1)}{\omega \varepsilon^{3n-1}(n-1)\mu_2^2 \lambda_{\bar{N}}^2},$
%$\frac{5\omega \sqrt{n-1}}{4 \varepsilon^{n-2}(\mu_2+\mu_2^2 \varepsilon \sqrt{n-1})}\}$.

%%%%%%%%%%%%%%%%%%%%%%%%%%%%%%%%%%%%%%%%%
\emph{Step 2: The convergence analysis of \eqref{eq orthogonal transformation final} to the origin.}

%Take the following functions
%\begin{align*}
%V_1=&\bar{\hat{\textbf{x}}}^T P_1 \bar{\hat{\textbf{x}}}\\
%V_2
%=& \tilde{\chi} _{1}^T P_2 \tilde{\chi} _{1} + \tilde{\chi} _{2}^T (P_2 \otimes I_{\bar{N}-1}) \tilde{\chi} _{2} \nonumber \\
%%%%%%%
%& +\frac{1}{2}\|  {\chi} _{1} +\sum_{i=1}^{n-2}  k_{i+1} {\chi} _{1}^{(i)} + {\chi} _{1}^{(n-1)}\|^2 \nonumber \\
%&+\frac{1}{2}\|  {\chi} _{2} +\sum_{i=1}^{n-2}  k_{i+1} {\chi} _{2}^{(i)} + {\chi} _{2}^{(n-1)}\|^2 \nonumber \\
%&+\frac{1}{2}\|\mu_1{\chi} _{1}^{(n-1)} + \mu_2 \eta_1\|^2 +\frac{1}{2}\|\mu_1{\chi} _{2}^{(n-1)} +\mu_2\eta_2\|^2 \nonumber
%\end{align*}
Take the following candidate Lyapunov function
\begin{align} \label{eq V}
V=&\tilde{\chi} _{1}^T P_1 \tilde{\chi} _{1} + \tilde{\chi} _{2}^T (P_1 \otimes I_{\bar{N}-1}) \tilde{\chi} _{2}+  \bar{\hat{\textbf{x}}}^T P_2 \bar{\hat{\textbf{x}}}\nonumber \\
%%%%%%
& +\frac{1}{2}\|k_1  {\chi} _{1} +\sum_{i=1}^{n-2}  k_{i+1} {\chi} _{1}^{(i)} + {\chi} _{1}^{(n-1)}\|^2 \nonumber\\
&+\frac{1}{2}\| k_1 {\chi} _{2} +\sum_{i=1}^{n-2}  k_{i+1} {\chi} _{2}^{(i)} + {\chi} _{2}^{(n-1)} + \mu \eta_2 \|^2
\end{align}
where
%$P_2 (\bm{\mathcal{L}} \otimes I_{\bar{N}}+ M) + (\bm{\mathcal{L}} \otimes I_{\bar{N}}+ M)^T P_2 =Q$, $P_2$ and $Q$ are symmetric positive-define matrices,
$P_1$ and $P_2$ are symmetric positive-define matrices such that $P_1A +A^T P_1 =-I_{n-1}$ and $P_2 (\bm{\mathcal{L}} \otimes I_{\bar{N}}+ M) + (\bm{\mathcal{L}} \otimes I_{\bar{N}}+ M)^T P_2 =Q$ hold, respectively (see Lemma \ref{lemma.cp}).

The derivative of $V$ along \eqref{eq orthogonal transformation final} is
\begin{align} \label{eq dV}
\dot{V}=&-\kappa_2\bar{\hat{\textbf{x}}}^T Q \bar{\hat{\textbf{x}}} -2\bar{\hat{\textbf{x}}}^TP_2 (1_{\bar{N}} \otimes \dot{\bar{\textbf{x}}})
  -\frac{k_1}{\varepsilon^{n-1}} ({\chi _{1}^T}r^T  + {\chi _{2}^T}R^T)h \nonumber \\
& -\varepsilon \|\tilde{\chi}\|^2 -\frac{\mu}{\varepsilon^{n-1}} \|\eta\|^2  -\frac{\mu}{\varepsilon^{n-1}} ({\eta_{1}^T}r^T  + {\eta_{2}^T}R^T)h \nonumber \\
& -\frac{1}{\varepsilon^{n-1}} \bigg(\sum_{l=1}^{n-2} (2 p_{l(n-1)} +k_{l+1}) \chi_1^{(l)} \nonumber \\
& \qquad \qquad  +(2 p_{(n-1)(n-1)} +1) \chi_1^{(n-1)}\bigg)^T (\eta_1 +r^T h) \nonumber \\
& -\frac{1}{\varepsilon^{n-1}} \bigg(\sum_{l=1}^{n-2} (2 p_{l(n-1)} +k_{l+1}) \chi_2^{(l)} \nonumber \\
& \qquad \qquad +(2 p_{(n-1)(n-1)} +1) \chi_2^{(n-1)} \bigg)^T (\eta_2 +R^T h) \nonumber \\
&  - \frac{k_1}{\varepsilon^{n-1}} \chi_2 \eta_2 + \kappa_1 k_1 \mu \chi_2 R^T \textbf{L} R \chi_2 + \kappa_1 \mu^2 \eta_2 R^T  \textbf{L} R \chi_2\nonumber\\
&+ \kappa_1 \mu \! \sum_{i=1}^{n-2}  k_{i+1} {\chi} _{2}^{(i)}\! R^T\!  \textbf{L} R \chi_2 \! + \!\kappa_1  \mu  \chi_2^{(n-1)}\! R^T\!  \textbf{L} R \chi_2
\end{align}
where $\tilde{\chi}=col(\tilde{\chi} _{1},\tilde{\chi} _{2})$.

Because $Q$ is a symmetric positive-define matrix, we have
\begin{equation} \label{eq dV Q_1}
- \kappa_2\bar{\hat{\textbf{x}}}^T Q \bar{\hat{\textbf{x}}} \leq - \kappa_2\lambda_{min} {(Q)}\|\bar{\hat{\textbf{x}}}\|^2.
\end{equation}

%By Young¡¯s inequality, we have
%\begin{align}\label{eq dV1 2}
%- 2\bar{\hat{\textbf{x}}}^T P_1 \dot{\hat{\textbf{x}}}^* \leq & 2 \|\bar{\hat{\textbf{x}}}\| \|P_1\| \|\dot{\hat{\textbf{x}}}^*\| \nonumber \\
%\leq & \varepsilon^{n+1} \|\bar{\hat{\textbf{x}}}\|^2 + \frac{\lambda_{max}^2(P_1)}{\varepsilon^{n+1}}\|\dot{\hat{\textbf{x}}}^*\|^2 \nonumber \\
%= &  \varepsilon^{n+1} \|\bar{\hat{\textbf{x}}}\|^2 + \frac{\lambda_{max}^2(P_1)}{\varepsilon^{n+1}}\|\dot{\bar{\textbf{x}}} + \dot{\textbf{x}}^*\|^2 \nonumber \\
%= &  \varepsilon^{n+1} \|\bar{\hat{\textbf{x}}}\|^2 + \frac{\varepsilon^2 \lambda_{max}^2(P_1)}{\varepsilon^{n+1}} \|\chi^{(1)} \|^2\nonumber \\
%= &  \varepsilon^{n+1} \|\bar{\hat{\textbf{x}}}\|^2 + \frac{ \lambda_{max}^2(P_1)}{\varepsilon^{n-1}} \|\chi^{(1)} \|^2
%\end{align}

%With \eqref{eq dV1}, \eqref{eq dV1 1} and \eqref{eq dV1 2}, we have
%\begin{align}
%\dot{V_1}\leq  -(\kappa_2 \lambda_{min}{(Q_1)}\! - \!\varepsilon^{n+1} ) \|\bar{\hat{\textbf{x}}} \|^2 \!+\! \frac{\lambda_{max}^2(P_1)}{\varepsilon^{n-1}}\|\chi^{(1)}\|^2
%\end{align}

According to the $\omega$-strongly monotonicity of $F(\textbf{x})$ (see Assumption \ref{Ass cost fc convex}), the Lipschitz continuity of $ F(\hat{\textbf{x}})$ (see Lemma \ref{lem.F}) and \eqref{eq orthogonal transformation}, we have
\begin{align}\label{eq dV convex}
 - (\chi _1^T{r^T}h +\chi _2^T{R^T}h)  % \nonumber\\
% & = - {(\textbf{x} - \textbf{x}^*)^T (F({\hat{\textbf{x}}}) - F(\hat{\textbf{x}}^*))} \nonumber\\
 & = - \bar{\textbf{x}} ^T (F({\hat{\textbf{x}}}) - F(1_{\bar{N}} \otimes {\textbf{x}}) + F({\textbf{x}}) - F(\hat{\textbf{x}}^*)) \nonumber\\
 %& = - \bar{\textbf{x}} ^T (F({\hat{\textbf{x}}}) - F(1_{\bar{N}}\otimes{\textbf{x}})\nonumber \\
% & ~~~\, + F(1_{\bar{N}}\otimes{\textbf{x}}) - F(\hat{\textbf{x}}^*)) \nonumber\\
% & \leq - \omega \|\bar{\textbf{x}}\|^2 + \| \bar{\textbf{x}} \| \| F({\hat{\textbf{x}}}) - F(1_{\bar{N}} \otimes {\textbf{x}}) \|  \nonumber\\
 & \leq - \omega \|\bar{\textbf{x}}\|^2 + \theta \| \bar{\textbf{x}} \| \| \hat{\textbf{x}} - 1_{\bar{N}} \otimes \textbf{x}\| \nonumber\\
 & \leq - \frac{\omega}{2} \| \chi \|^2 + \frac{\theta^2}{2\omega} \|\bar{\hat{\textbf{x}}} \|^2.
\end{align}

It results from the orthogonal transformation \eqref{eq orthogonal transformation} and the Lipschitz continuity of $F(\hat{\textbf{x}})$ that
\begin{align}\label{eq dV Lipschitz 1}
-\frac{\mu}{\varepsilon^{n-1}} ({\eta_{1}^T}r^T  + {\eta_{2}^T}R^T)h
 %=&  - k{{\bar v}^T}h \nonumber\\
 =&  -\frac{\mu}{\varepsilon^{n-1}} {\bar{\textbf{y}}^T (F({\hat{\textbf{x}}}) - F(\hat{\textbf{x}}^*))} \nonumber\\
  \le &  \frac{\mu \theta}{\varepsilon^{n-1}} \|\bar{\textbf{y}} \|  \|\bar{\hat{\textbf{x}}}\| \nonumber\\
 \leq &  \frac{1}{4\varepsilon^{n-1}}\|\eta_2 \|^2 +  \frac{\theta^2 \mu^2}{\varepsilon^{n-1}}    \|\bar{\hat{\textbf{x}}}\|^2
\end{align}
and
\begin{align}\label{eq dV Lipschitz 2}
\,&-\frac{1}{\varepsilon^{n-1}} \bigg(\sum_{l=1}^{n-2} (2 p_{l(n-1)} +k_{l+1}) \chi_1^{(l)} \nonumber \\
& \qquad \qquad  +(2 p_{(n-1)(n-1)} + 1  ) \chi_1^{(n-1)}\bigg)^T r^T h \nonumber \\
& -\frac{1}{\varepsilon^{n-1}} \bigg(\sum_{l=1}^{n-2} (2 p_{l(n-1)} +k_{l+1}) \chi_2^{(l)} \nonumber \\
& \qquad \qquad +(2 p_{(n-1)(n-1)} +1 ) \chi_2^{(n-1)} \bigg)^T R^T h \nonumber \\
& \leq \frac{\theta}{\varepsilon^{n-1}} \bigg(\sum_{l=1}^{n-2} |2 p_{l(n-1)} +k_{l+1}| \|\tilde{\textbf{x}}^{(l)}\| \nonumber \\
& \qquad \qquad  +|2 p_{(n-1)(n-1)} +1 | \| \tilde{\textbf{x}}^{(n-1)}\| \bigg)^T \| \bar{\hat{\textbf{x}}} \|  \nonumber \\
& \leq \frac{\theta}{\varepsilon^{n-1}} \bigg( \frac{1}{2} \big(\sum_{l=1}^{n-2} (2 p_{l(n-1)} +k_{l+1})^2 \|\chi^{(l)}\|^2 \nonumber \\
& \qquad \qquad  + (2 p_{(n-1)(n-1)} + 1 +)^2 \|\chi^{(n-1)}\|^2 \big)\nonumber\\
& \qquad \qquad + \frac{n-1}{2} \|\bar{\hat{\textbf{x}}} \|^2\bigg).
\end{align}

Based on Young's inequality, we have
\begin{subequations} \label{eq dV Young}
\begin{align}
%%%%%%%%%%%%%%%%%%%%%%%%%%%%%%%%%%%%%%%%%%%%%%%%%%%%%
&-2\bar{\hat{\textbf{x}}}^TP_2 (1_{\bar{N}} \otimes \dot{\bar{\textbf{x}}})
\leq 3 \varepsilon \lambda_{max}^2(P_2)  \|\bar{\hat{\textbf{x}}} \|^2 + \frac{1}{6\varepsilon}\|\dot{\bar{\textbf{x}}}\|^2 \nonumber \\
& \qquad \qquad \qquad \qquad  \leq 6 \varepsilon \lambda_{max}^2(P_2)  \|\bar{\hat{\textbf{x}}} \|^2 + \frac{\varepsilon}{6} \|\chi^{(1)}\|^2 \\
%%%%%%%%%%%%%%%%%%%%%%%%%%%%%%%%%%%%%%%%%%%%%%%%%%%%%
&-\frac{1}{\varepsilon^{n-1}} \bigg(\sum_{l=1}^{n-2} (2 p_{l(n-1)} +k_{l+1}) \chi_1^{(l)} \nonumber \\
& \qquad \qquad  +(2 p_{(n-1)(n-1)} +1 ) \chi_1^{(n-1)} \bigg)^T \eta_1 \nonumber \\
& -\frac{1}{\varepsilon^{n-1}} \bigg(\sum_{l=1}^{n-2} (2 p_{l(n-1)} +k_{l+1}) \chi_2^{(l)} \nonumber \\
& \qquad \qquad +(2 p_{(n-1)(n-1)} +1 ) \chi_2^{(n-1)} \bigg)^T \eta_2 \nonumber \\
& \leq \frac{1}{\varepsilon^{n-1}} \bigg(\sum_{l=1}^{n-2} |2 p_{l(n-1)} +k_{l+1}| (\|{\chi}_1^{(l)}\| \|\eta_1\| + \|{\chi}_2^{(l)}\| \|\eta_2\|)\nonumber \\
& \qquad \qquad +|2 p_{(n-1)(n-1)} +1 | (\|{\chi}_1^{(n-1)}\| \|\eta_1\| \nonumber \\
& \qquad \qquad + \|{\chi}_2^{(n-1)}\| \|\eta_2\|) \bigg)\nonumber \\
& \leq \frac{1}{\varepsilon^{n-1}} \bigg(\frac{n-1}{2} \|\eta\|^2 +\frac{1}{2} \big(\sum_{l=1}^{n-2} (2 p_{l(n-1)} +k_{l+1})^2 \|{\chi}^{(l)}||^2 \nonumber \\
& \qquad \qquad  +(2 p_{(n-1)(n-1)} +1)^2 \|{\chi}^{(n-1)} \|^2 \big) \bigg)  \\
%%%%%%%%%%%%%%%%%%%%%%%%%%%%%%%%%%%%%%%%%%%%%%%%%%%%%%%
&- \frac{k_1}{\varepsilon^{n-1}} \chi_2 \eta_2
\leq \frac{\omega k_1}{4\varepsilon^{n-1}} \|\chi_2 \|^2 + \frac{k_1}{\omega\varepsilon^{n-1}} \|\eta_2\|^2  \\
%%%%%%%%%%%%%%%%%%%%%%%%%%%%%%%%%%%%%%%%%%%%%%%%%%%%%%%
&\kappa_1k_1 \mu \chi_2 R^T  \textbf{L} R \chi_2 \leq \kappa_1k_1 \mu \|\textbf{L} \| \|\chi_2\|^2  \\
%%%%%%%%%%%%%%%%%%%%%%%%%%%%%%%%%%%%%%%%%%%%%%%%%%%%%%
& \kappa_1 \mu \sum_{i=1}^{n-2}  k_{i+1} {\chi} _{2}^{(i)} R^T  \textbf{L} R \chi_2
\leq \frac{\varepsilon \kappa_1\| \textbf{L} \|^2 \mu^2}{6} \sum_{i=1}^{n-2} ( k_{i+1})^2 \|{\chi} _{2}^{(i)}\|^2   \nonumber \\
&\qquad \qquad\qquad\qquad\qquad \qquad + \frac{3(n-2)}{2\varepsilon} \kappa_1 \| \chi_2 \|^2  \\
%%%%%%%%%%%%%%%%%%%%%%%%%%%%%%%%%%%%%%%%%%%%%%%
& \kappa_1 \mu \chi_2^{(n-1)} R^T  \textbf{L} R \chi_2  \leq \frac{\varepsilon \kappa_1\| \textbf{L} \|^2 \mu^2}{6} \|{\chi} _{2}^{(n-1)}\|^2
 + \frac{3}{2\varepsilon} \kappa_1 \| \chi_2 \|^2 \\
%%%%%%%%%%%%%%%%%%%%%%%%%%%%%%%%%%%%%%%%%%%%%%%%%%%%%%
 & \kappa_1 \mu^2 \eta_2 R^T  \textbf{L} R \chi_2  \leq \kappa_1 \frac{\mu^2}{2} \| \textbf{L}\|^2 \|\eta_2 \|^2 +  \kappa_1 \frac{\mu^2}{2  } \|\chi_2 \|^2.
\end{align}
\end{subequations}

Combining \eqref{eq dV}-\eqref{eq dV Young}, we have
\begin{align}\label{eq dotV negative}
 \dot{V} \leq & -\bigg{(}\frac{2\varepsilon}{3} - \frac{1}{\varepsilon^{n-1}}(\frac{\theta \bar{a}_1}{2} + \frac{\bar{a}_1}{2}) \bigg{)}\|\tilde{\chi} \|^2 \nonumber \\
 &-\bigg{(}\frac{\omega k_1}{4\varepsilon^{n-1}} - \kappa_1(\frac{3n-3}{2\varepsilon} + \frac{\mu^2}{2} + k_1 \mu \|\textbf{L} \|)\bigg{)}\|\chi \|^2 \nonumber \\
 &-\bigg{(}\frac{1}{\varepsilon^{n-1}} (\mu- \frac{ 2n -1 }{4} - \frac{k_1}{\omega}) - \frac{1}{2}\mu^2 \kappa_1\| \textbf{L}\|^2\bigg{)}\|\eta_2 \|^2 \nonumber \\
 &-\bigg{(}\kappa_2\lambda_{min} {(Q)}  -\frac{\theta^2 k_1 + 2 \omega \mu^2\theta^2 + \omega\theta(n-1)}{2\omega\varepsilon^{n-1}} \nonumber \\
 & \qquad -6 \varepsilon \lambda_{max}^2(P_2) \bigg{)}\|\bar{\hat{\textbf{x}}}\|^2.
\end{align}

\eqref{eq dotV negative} shows that $\dot V$ is negative definite. Besides, the Lyapunov function $V$ and its derivative are quadratic, which indicates that
\eqref{eq orthogonal transformation final} is globally exponentially stable, i.e., the high-order player \eqref{eq sys} under the algorithm \eqref{eq algorithm} globally exponentially converges to the Nash equilibrium of the multi-cluster game \eqref{eq multi-cluster game}.
\hfill $\Box$

\begin{remark}
%Although the dynamics of the participants are high-order, Theorem \ref{the.1} assures the high-order player \eqref{eq sys} exponentially converges to the Nash equilibrium of the multi-cluster game \eqref{eq multi-cluster game}.
The algorithm \eqref{eq algorithm} is exponentially convergent, which is different form the asymptotically convergent algorithms in \cite{ZhangLWJ19,deng2021distributedEL,bianchiG2021,RomanoP20,ZENG201920}.
 Furthermore, Compared with the algorithms in \cite{Ye2019AUnifiedStrategy} and \cite{pang2020gradient}, the algorithm \eqref{eq algorithm} converges to the exact Nash equilibrium rather than the neighborhood of the Nash equilibrium.
%In contrast to many existing multi-cluster game algorithms that require every player to have access to decisions of players in the same and different clusters (see \cite{YE2018266}, \cite{Ye2019AUnifiedStrategy}, \cite{pang2020gradient}, \cite{ZENG201920}), player $i$ in cluster $j$ only exchanges necessary information with its neighbors in cluster $j$, but does not share its decision with players in other clusters. Besides, the cost functions and gradients of players are not shared with their neighbors, which signifies that algorithm \eqref{eq algorithm} is conducive to privacy protection.
%all players only exchange with their neighbors.
\end{remark}

%\begin{remark}
%Theorem \ref{the.1} implies that under the algorithm \eqref{alg}, the high-order system \eqref{sys} can autonomously complete the resource allocation task \eqref{pro}.
%%Besides, different from the results in \cite{Xiao06,Lakshmanan08,Cherukuri15,Wei13}, the network resource information is not used in our algorithm.
%Moreover, the algorithm \eqref{alg} can also solve the RAPs considered in \cite{Xiao06,Lakshmanan08,DengLY18} by letting $n=1$ or $2$.
%Besides, by the algorithm \eqref{alg}, the network resource information, the gradients, the cost functions and the decisions of agents are not shared among networks, contrast to some existing distributed resource allocation algorithms such as \cite{Xiao06,Lakshmanan08,Wei13,Cherukuri15}.
%%In order to obtain the which may require an additional algorithm to get the information.
%%{\color{blue}Additionally, }
%\end{remark}

%\subsection{Application in Power Grid}

\begin{figure}[htbp]
\setcaptionwidth{10in}
\centering
%\subfigure[Communication topology of one group company.]{\includegraphics[width=6cm,height=3cm]{smallonecluster}}
%\subfigure[Communication topology of all companies.]
{\includegraphics[width=8.5cm,height=2.5cm]{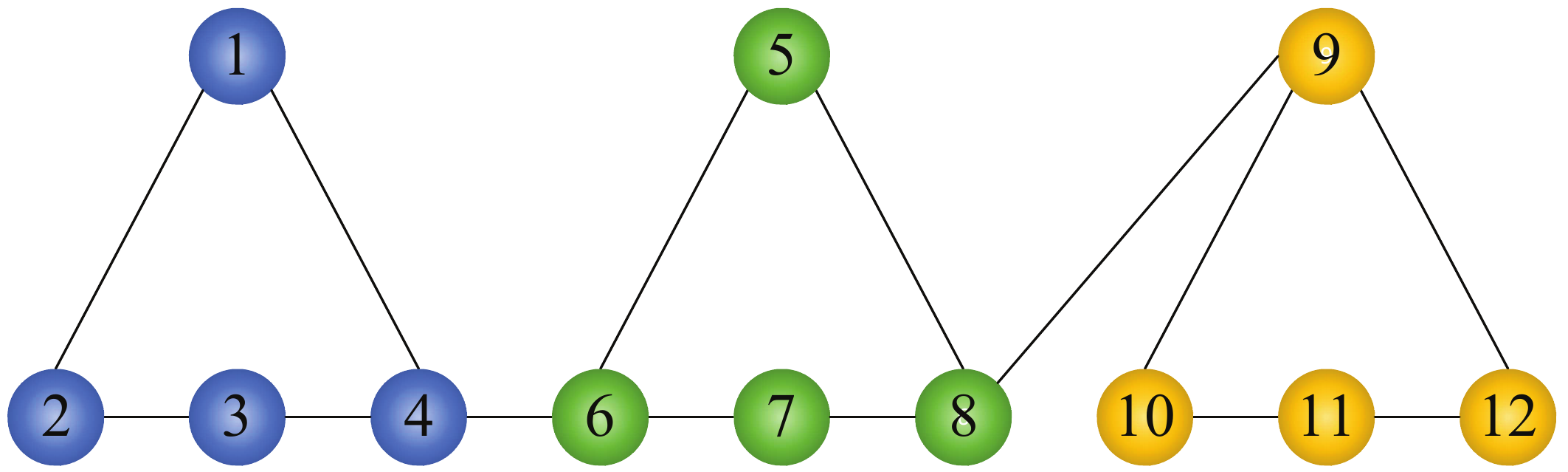}}
\caption{The communication topology.}
\label{fig communication graph}
\end{figure}
\section{Numerical Examples}\label{sec.s}

In this section, a numerical example is presented to illustrate the algorithm \eqref{eq algorithm}.

Consider a multi-cluster game with $N$ clusters, where cluster $j \in \{ 1,\dots,N\}$ is composed of $n_j$ players. The clusters compete with each other for their own benefits, and the players in the same cluster cooperate with each other. Cluster $j \in \{ 1,\dots,N\}$ faces the following multi-cluster game:
\begin{equation*} %\label{eq simu multi-cluster game}
\begin{split}
&\min_{\textbf{x}^j \in \mathbb{R}^{qn_j}}f^j \left(\textbf{x}^j,\textbf{x}^{-j} \right)\\
&\,x_i^j = x_k^j, \, \forall i,k \in \{ 1, \cdots, n_j \}
\end{split}
\end{equation*}
where $f^j ( \textbf{x}^{j},\textbf{x}^{-j} )=\sum_{i=1}^{n_j} f_i^j ( x_i^{j},\textbf{x}^{-j})$ is the cost function of cluster $j$, $f_i^j ( x_i^{j},\textbf{x}^{-j})$ is the cost function of player $i$ in cluster $j$, $x_i^j$ is the decision of player $i$ in cluster $j$,
$\textbf{x}^j=col ( x^j_1 , \dots,  x^j_{n_j} )$,
$\textbf{x}^{-j}=col (\textbf{x}^1 ,\textbf{x}^2, \dots, \textbf{x}^{j-1}, \textbf{x}^{j+1}, \dots, \textbf{x}^N)$.
Here, we consider $N=3$, $n_j=4$, $\forall j \in \{ 1,\dots,N\}$. Particularly, the cost functions of all players are expressed as follows:
\begin{subequations}
\begin{align*}
f_1^1 ( x_1^1,\textbf{x}^{-1} )=&\frac{3.4 (x_1^1)^2}{6.2 \sqrt{5.2 (x_1^1)^2 +27}}+1.9 (x_1^1)^2 -60x_1^1 \\
f_2^1 ( x_2^1,\textbf{x}^{-1} )=&\frac{2.3 (x_2^1)^2}{1.5 \sqrt{2.8 (x_2^1)^2 +50}}+3.3 (x_2^1)^2 -38x_2^1 \\
f_3^1 ( x_3^1,\textbf{x}^{-1} )=&\frac{2.1 (x_3^1)^2}{2.9 \sqrt{3.4 (x_3^1)^2 +42}}+2.6 (x_3^1)^2 -75x_3^1 \\
f_4^1 ( x_4^1,\textbf{x}^{-1} )=&\frac{4.1 (x_4^1)^2}{5.6 \sqrt{4.4 (x_4^1)^2 +47}}+1.6 (x_4^1)^2 -65x_4^1 \\
&+ \hat{x}_{22}^{14}x_4^1 \\
f_1^2 ( x_1^2,\textbf{x}^{-2} )=&\frac{1.8 (x_1^2)^2}{2ln(2.3(x_1^2)^2+80)}+ 3(x_1^2)^2 - 50 x_1^2\\
f_2^2 ( x_2^2,\textbf{x}^{-2} )=&\frac{4.2 (x_2^2)^2}{4.8ln(6.2(x_2^2)^2+20)}+ 2.4(x_2^2)^2 - 40 x_2^2 \\
&+ \hat{x}_{14}^{22}x_2^2\\
f_3^2 ( x_3^2,\textbf{x}^{-2} )=&\frac{2.3 (x_3^2)^2}{4.3ln(5.7(x_3^2)^2+38)}+ (x_3^2)^2 - 55 x_3^2\\
f_4^2 ( x_4^2,\textbf{x}^{-2} )=&\frac{3.3 (x_4^2)^2}{2.5ln(6.1(x_4^2)^2+48)}+ 2.8(x_4^2)^2 - 42 x_4^2 \\
&+ \hat{x}_{31}^{24}x_4^2\\
f_1^3 ( x_1^3,\textbf{x}^{-3} )=&4.3 (x_1^3)^2 -43 x_1^3 +20 +\hat{x}_{24}^{31}x_1^3 \\
f_2^3 ( x_2^3,\textbf{x}^{-3} )=&2.5 (x_2^3)^2 -34 x_2^3 +45 \\
f_3^3 ( x_3^3,\textbf{x}^{-3} )=&3.7 (x_3^3)^2 -36 x_3^3 +12 \\
f_4^3 ( x_4^3,\textbf{x}^{-3} )=&2.7 (x_4^3)^2 -40 x_4^3 +24.
\end{align*}
\end{subequations}

The dynamics of player $i$ in cluster $j$ are $x_i^{j(4)}=u_i^j$. %, where $u_i^j$ is the control input.
The communication topology among players is depicted as Fig. \ref{fig communication graph}. The algorithm parameters are chosen as $k_1=1$, $k_2=2$, $k_3=1$, $\varepsilon = 3.71$, $\kappa_1=0.05$ and $\kappa_2 = 386$.
%The initial values of $\textbf{x}$, $\bm{\lambda}$, $\bm{\mu}$, $\hat{\textbf{x}}$ are zeros and $\textbf{x}=[25,55,35,45,25,20,35,50,45,30,25,35]^T$.

\begin{figure}[htbp]
\setcaptionwidth{10in}
\centering
%\subfigure[Communication topology of one group company.]{\includegraphics[width=6cm,height=3cm]{smallonecluster}}
%\subfigure[Communication topology of all companies.]
{\includegraphics[width=9.2cm,height=7cm]{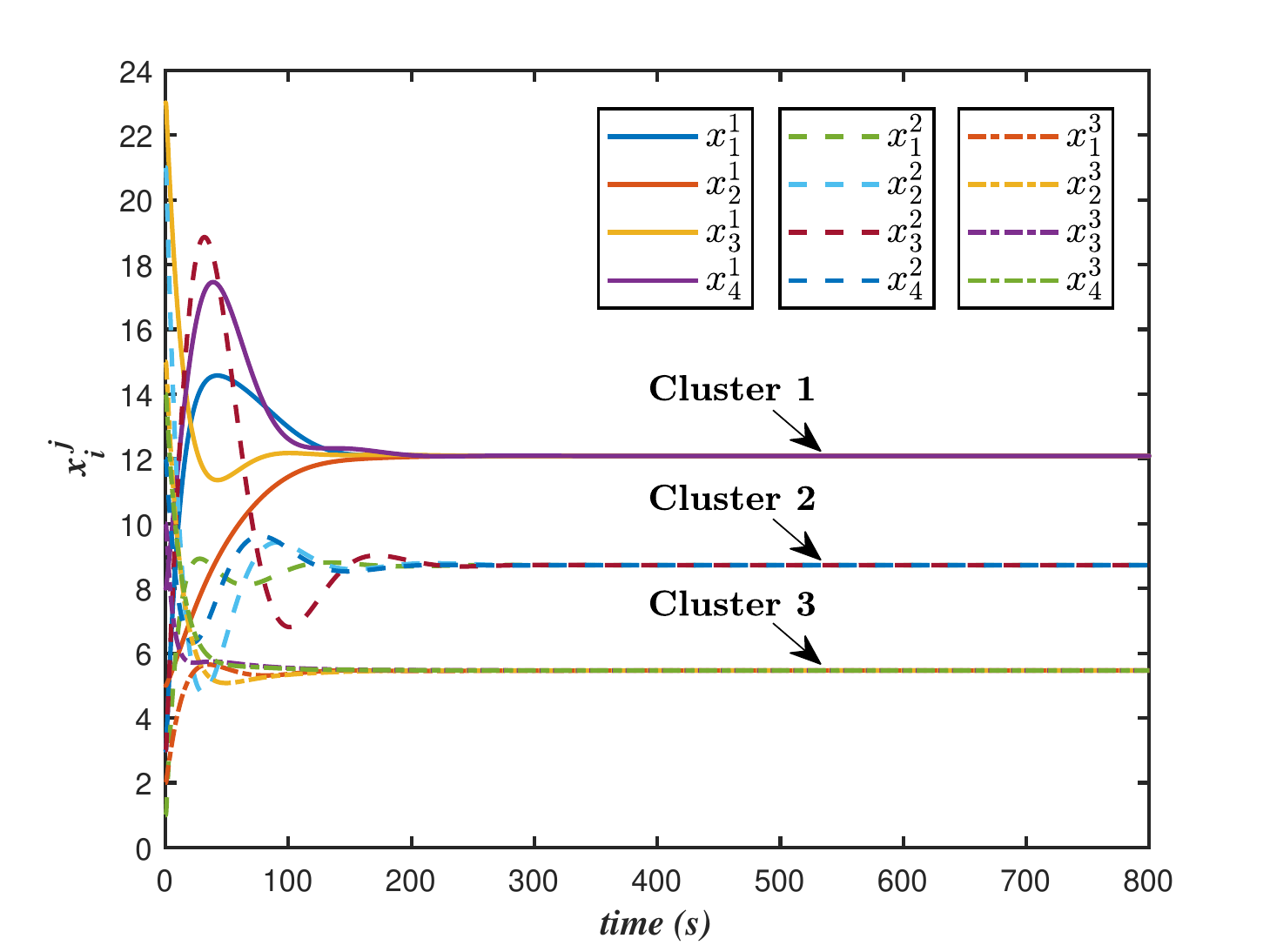}}
\caption{Evolutions of $x_i^j$.}
\label{fig results}
\end{figure}

The simulation results are presented in Fig. \ref{fig results}, where the solid lines, the dotted lines and the dot-dash lines are the evolutions of outputs of clusters 1, 2 and 3, respectively.
%, where the solid and dot-dash lines are the evolutions of strategy profile and the variational GNE of the multi-cluster game \eqref{eq simu multi-cluster game}.
As shown in Fig. \ref{fig results}, the decisions of players in the same cluster  reach a common strategy, and the decisions of all players converge to the Nash equilibrium under algorithm \eqref{eq algorithm}.
These simulation results verify the effectiveness of our method.

%\begin{figure}[htbp]
%%\setcaptionwidth{10in}
%\centering
%%\vspace{-2.5cm}%ͼƬÓëÒ³ÃæÉÏ·½µÄ¾àÀë
%\setlength{\abovecaptionskip}{-2.5pt}% ͼÓëÖ÷±êÌâµÄ¾àÀë
%%\subfigtopskip=-1pt %ÉèÖÃ×ÓͼÓëÉÏÃæÕýÎÄ»ò±ðµÄÄÚÈݵľàÀë
%%	\subfigbottomskip=-3pt %ÉèÖõڶþÐÐ×ÓͼÓëµÚÒ»ÐÐ×ÓͼµÄ¾àÀ룬¼´ÏÂÃæµÄÍ·ÓëÉÏÃæµÄ½ÅµÄ¾àÀë
%\subfigcapskip=-1pt %ÉèÖÃ×ÓͼÓë×Ó±êÌâÖ®¼äµÄ¾àÀë
%\subfigure[]{\includegraphics[width=9.2cm,height=7cm]{x^1}}
%\subfigure[]{\includegraphics[width=9.2cm,height=7cm]{x^2}}\hspace{-10mm}
%\subfigure[]{\includegraphics[width=9.2cm,height=7cm]{x^3}}\hspace{-10mm}
%\caption{Evolutions of $x_i^j$.}
%\label{fig results}
%\end{figure}

\section{Conclusions} \label{sec.conclusion}
This paper has investigated the multi-cluster games of high-order multi-agent systems. To seek the Nash equilibrium of the multi-cluster game, we have designed a distributed algorithm via gradient descent and state
feedback. In the algorithm, a distributed estimator has been employed such that players can estimate the decisions of other players.
In comparison with other results for multi-cluster games, players only need to share some information with their
neighbors by our algorithm.
%cost functions and gradients of players are not exchanged among networks in our algorithm.
Besides, we have analyzed the convergence of the algorithm. Under the algorithm, all high-order players exponentially converge to the exact Nash equilibrium of the multi-cluster game. Finally, a numerical example has illustrated the result.

%This paper  has investigated RAPs of high-order multi-agent systems with coupling constraints.
%In the problem, the dynamics of agents are  high-order integrators and the available local resources make up of the network resource.
%%In order to ensure the multi-agent system achieves the optimal resource allocation,
%To solve the problem, a distributed continuous-time algorithm has been designed for the high-order agents.
%With the algorithm, the agents do not share their decisions, gradients and cost functions with their neighbors, and do not need to know the network resource information.
%Besides, the convergence of the algorithm has been analyzed.
%By the algorithm, the high-order agents globally exponentially converge the optimal resource allocation.
%Furthermore, the algorithm has been illustrated by two numerical examples.
%{\color{blue}In fact, many challenging problems remain to be done, such as RAPs of high-order systems over directed graphs, RAPs with  inequality resource constraints, and RAPs with general convex constraints.}

\appendices

%\section*{Acknowledgement}
%
%This work was supported by the National Key Research and Development Program of China (2016YFB0901902), NSFC (61333001, 61573344) and the China Postdoctoral Science Foundation (No. 2016M591272).

\ifCLASSOPTIONcaptionsoff
  \newpage
\fi

\footnotesize

\bibliographystyle{IEEEtran}
\bibliography{ref}           % and a bib file to produce the

\begin{thebibliography}{10}
\providecommand{\url}[1]{#1}
\csname url@samestyle\endcsname
\providecommand{\newblock}{\relax}
\providecommand{\bibinfo}[2]{#2}
\providecommand{\BIBentrySTDinterwordspacing}{\spaceskip=0pt\relax}
\providecommand{\BIBentryALTinterwordstretchfactor}{4}
\providecommand{\BIBentryALTinterwordspacing}{\spaceskip=\fontdimen2\font plus
\BIBentryALTinterwordstretchfactor\fontdimen3\font minus
  \fontdimen4\font\relax}
\providecommand{\BIBforeignlanguage}[2]{{%
\expandafter\ifx\csname l@#1\endcsname\relax
\typeout{** WARNING: IEEEtran.bst: No hyphenation pattern has been}%
\typeout{** loaded for the language `#1'. Using the pattern for}%
\typeout{** the default language instead.}%
\else
\language=\csname l@#1\endcsname
\fi
#2}}
\providecommand{\BIBdecl}{\relax}
\BIBdecl

\bibitem{Gharesifard2016Price}
B.~Gharesifard, T.~Basar, and A.~D. Dominguez-Garcia, ``Price-based coordinated
  aggregation of networked distributed energy resources,'' \emph{IEEE
  Transactions on Automatic Control}, vol.~61, no.~10, pp. 2936--2946, Oct.
  2016.

\bibitem{ghaderi2014opinion}
J.~Ghaderi and R.~Srikant, ``Opinion dynamics in social networks with stubborn
  agents: Equilibrium and convergence rate,'' \emph{Automatica}, vol.~50,
  no.~12, pp. 3209--3215, 2014.

\bibitem{Ram2010ParameterEstimation}
S.~S. Ram, V.~V. Veeravalli, and A.~Nedi\'{c}, ``Distributed and recursive
  parameter estimation in parametrized linear state-space models,'' \emph{IEEE
  Transactions on Automatic Control}, vol.~55, no.~2, pp. 488--492, Feb. 2010.

\bibitem{cao2020}
M.~Cao, ``Merging game theory and control theory in the era of {AI} and
  autonomy,'' \emph{National Science Review}, vol.~7, no.~7, pp. 1122--1124,
  2020.

\bibitem{Yuan2021}
D.~Yuan, D.~W.~C. Ho, and S.~Xu, ``Stochastic strongly convex optimization via
  distributed epoch stochastic gradient algorithm,'' \emph{IEEE Transactions on
  Neural Networks and Learning Systems}, pp. 1--14, 2020, to be published.

\bibitem{Lou16}
Y.~Lou, Y.~Hong, L.~Xie, G.~Shi, and K.~H. Johansson, ``Nash equilibrium
  computation in subnetwork zero-sum games with switching communications,''
  \emph{IEEE Transactions on Automatic Control}, vol.~61, no.~10, pp.
  2920--2935, Oct. 2016.

\bibitem{liang2019exponential}
S.~Liang, L.~Y. Wang, and G.~Yin, ``Exponential convergence of distributed
  primal--dual convex optimization algorithm without strong convexity,''
  \emph{Automatica}, vol. 105, pp. 298--306, 2019.

\bibitem{yang2017distributed}
S.~Yang, Q.~Liu, and J.~Wang, ``Distributed optimization based on a multiagent
  system in the presence of communication delays,'' \emph{IEEE Transactions on
  Systems, Man, and Cybernetics: Systems}, vol.~47, no.~5, pp. 717--728, May
  2017.

\bibitem{liy2019}
R.~Li and G.-H. Yang, ``Consensus control of a class of uncertain nonlinear
  multiagent systems via gradient-based algorithms,'' \emph{IEEE transactions
  on cybernetics}, vol.~49, no.~6, pp. 2085--2094, Jun. 2019.

\bibitem{he2019continuous}
X.~He, T.~Huang, J.~Yu, C.~Li, and Y.~Zhang, ``A continuous-time algorithm for
  distributed optimization based on multiagent networks,'' \emph{IEEE
  Transactions on Systems, Man, and Cybernetics: Systems}, vol.~49, no.~12, pp.
  2700--2709, Dec. 2019.

\bibitem{kia2015distributed}
S.~S. Kia, J.~Cort{\'e}s, and S.~Mart{\'\i}nez, ``Distributed convex
  optimization via continuous-time coordination algorithms with discrete-time
  communication,'' \emph{Automatica}, vol.~55, pp. 254--264, 2015.

\bibitem{Zhang17}
Y.~Zhang, Z.~Deng, and Y.~Hong, ``Distributed optimal coordination for multiple
  heterogeneous {E}uler-{L}agrangian systems,'' \emph{Automatica}, vol.~79, pp.
  207--213, May 2017.

\bibitem{ZhangLWJ19}
Y.~{Zhang}, S.~{Liang}, X.~{Wang}, and H.~{Ji}, ``Distributed {N}ash
  equilibrium seeking for aggregative games with nonlinear dynamics under
  external disturbances,'' \emph{IEEE Transactions on Cybernetics}, vol.~50,
  no.~12, pp. 4876--4885, Dec. 2020.

\bibitem{deng2021distributedEL}
Z.~Deng, ``Distributed algorithm design for aggregative games of
  {E}uler-{L}agrange systems and its application to smart grids,'' \emph{IEEE
  Transactions on Cybernetics}, 2021, to be published.

\bibitem{bianchiG2021}
M.~Bianchi and S.~Grammatico, ``Continuous-time fully distributed generalized
  {Nash} equilibrium seeking for multi-integrator agents,'' \emph{Automatica},
  vol. 129, p. 109660, 2021.

\bibitem{RomanoP20}
A.~R. {Romano} and L.~{Pavel}, ``Dynamic {NE} seeking for multi-integrator
  networked agents with disturbance rejection,'' \emph{IEEE Transactions on
  Control of Network Systems}, vol.~7, no.~1, pp. 129--139, Mar. 2020.

\bibitem{shehory1998methods}
O.~Shehory and S.~Kraus, ``Methods for task allocation via agent coalition
  formation,'' \emph{Artificial intelligence}, vol. 101, no. 1-2, pp. 165--200,
  1998.

\bibitem{peng2009coexistence}
T.~J. Peng, A., and M.~Bourne, ``The coexistence of competition and cooperation
  between networks: implications from two taiwanese healthcare networks,''
  \emph{British Journal of Management}, vol.~20, no.~3, pp. 377--400, 2009.

\bibitem{YE2018266}
M.~{Ye}, G.~{Hu}, and F.~L. {Lewis}, ``{N}ash equilibrium seeking for
  {N}-coalition noncooperative games,'' \emph{Automatica}, vol.~95, pp.
  266--272, 2018.

\bibitem{ZENG201920}
X.~{Zeng}, J.~{Chen}, S.~{Liang}, and Y.~{Hong}, ``Generalized {N}ash
  equilibrium seeking strategy for distributed nonsmooth multi-cluster game,''
  \emph{Automatica}, vol. 103, pp. 20--26, 2019.

\bibitem{Ye2019AUnifiedStrategy}
M.~{Ye}, G.~{Hu}, F.~L. {Lewis}, and L.~{Xie}, ``A unified strategy for
  solution seeking in graphical {N}-coalition noncooperative games,''
  \emph{IEEE Transactions on Automatic Control}, vol.~64, no.~11, pp.
  4645--4652, Nov. 2019.

\bibitem{meng2020linear}
M.~Meng and X.~Li, ``On the linear convergence of distributed {N}ash
  equilibrium seeking for multi-cluster games under partial-decision
  information,'' \emph{arXiv preprint arXiv:2005.06923}, 2020.

\bibitem{pang2020gradient}
Y.~Pang and G.~Hu, ``Gradient-free {N}ash equilibrium seeking in {N}-cluster
  games with uncoordinated constant step-sizes,'' \emph{arXiv preprint
  arXiv:2008.13088}, 2020.

\bibitem{Kim12cyber}
K.~D. Kim and P.~R. Kumar, ``Cyber-physical systems: A perspective at the
  centennial,'' \emph{Proceedings of the IEEE}, vol. 100, pp. 1287--1308, May.
  2012.

\bibitem{Zhang18}
X.~Zhang, A.~Papachristodoulou, and N.~Li, ``Distributed control for reaching
  optimal steady state in network systems: An optimization approach,''
  \emph{IEEE Transactions on Automatic Control}, vol.~63, no.~3, pp. 864--871,
  Mar. 2018.

\bibitem{deng2019distributedre}
Z.~{Deng}, ``Distributed algorithm design for resource allocation problems of
  second-order multiagent systems over weight-balanced digraphs,'' \emph{IEEE
  Transactions on Systems, Man, and Cybernetics: Systems}, vol.~51, no.~6, pp.
  3512--3521, June. 2021.

\bibitem{Godsil01graph}
C.~D. Godsil and G.~Royle, \emph{Algebraic Graph Theory}.\hskip 1em plus 0.5em
  minus 0.4em\relax New York: Springer, 2001.

\bibitem{facchinei2003finite}
F.~Facchinei and J.-S. Pang, \emph{Finite-Dimensional Variational Inequalities
  and Complementarity Problems}.\hskip 1em plus 0.5em minus 0.4em\relax
  Springer Science \& Business Media, 2003.

\bibitem{holt2004nash}
C.~A. Holt and A.~E. Roth, ``The {N}ash equilibrium: A perspective,''
  \emph{Proceedings of the National Academy of Sciences}, vol. 101, no.~12, pp.
  3999--4002, 2004.

\bibitem{deng2018distributedgeneralized}
Z.~Deng and X.~Nian, ``Distributed generalized {N}ash equilibrium seeking
  algorithm design for aggregative games over weight-balanced digraphs,''
  \emph{IEEE Transactions on Neural Networks and Learning Systems}, vol.~30,
  no.~3, pp. 695--706, Mar. 2018.

\bibitem{facchinei2010generalized}
F.~Facchinei and C.~Kanzow, ``Generalized {N}ash equilibrium problems,''
  \emph{Annals of Operations Research}, vol. 175, no.~1, pp. 177--211, 2010.

\bibitem{Chen99}
C.-T. Chen, \emph{Linear System Theory and Design}, 3rd~ed.\hskip 1em plus
  0.5em minus 0.4em\relax New York: Oxford University Press, 1999.

\end{thebibliography}

\end{document}